\documentclass[10pt, journal]{IEEEtran}

\usepackage{epsfig}
\usepackage{amsmath}
\usepackage[T1]{fontenc}
\usepackage{times}
\usepackage{amsmath,bm}
\usepackage{tabularx}
\usepackage{makecell}
\usepackage{comment}
\usepackage{lipsum}
\usepackage{multirow}
\usepackage{array}
\usepackage{cuted}
\usepackage{caption}
\usepackage{subcaption}
\IEEEoverridecommandlockouts

\fontfamily{ptm}\selectfont

\begin{document}

\title{Modeling, Simulation and Fairness Analysis of Wi-Fi and Unlicensed LTE Coexistence}

\author{\IEEEauthorblockN{Morteza Mehrnoush, Rohan Patidar, Sumit Roy, and Thomas Henderson}\\
\IEEEauthorblockA{University of Washington, Seattle, WA-98195 \\
Email: \{mortezam, rpatidar, sroy, tomhend\}@uw.edu} \thanks{This work was supported in part by the National Science Foundation (NSF) under Award 1617153.}}

\maketitle


\begin{abstract}
Coexistence of small-cell LTE and Wi-Fi networks in unlicensed bands at $5$ GHz is a topic of active interest, primarily driven by industry groups affiliated with the two (cellular and Wi-Fi) segments. A notable alternative to the 3GPP Rel. 13 defined LTE-Licensed Assisted Access (LTE-LAA) mechanism for coexistence is the unlicensed LTE (LTE-U) Forum \cite{lteuforum} that prescribed Carrier Sense Adaptive Transmission (CSAT) whereby LTE utilizes the unlicensed band as a supplemental downlink unlicensed carrier (to enhance downlink data rate) to normal operation using licensed spectrum. In this work, we provide a new analytical model for performance analysis of unlicensed LTE with fixed duty cycling (LTE-DC) in coexistence with Wi-Fi. Further, the analytical results are cross-validated with ns-3 (www.nsnam.org) based simulation results using a newly developed coexistence stack. Thereafter, notions of {\em fair coexistence} are investigated that can be achieved by tuning the LTE duty cycle. The results show that as the number of Wi-Fi nodes increases, the Wi-Fi network in coexistence with LTE-DC with 0.5 duty cycling achieves a higher throughput than with an identical Wi-Fi network.

\end{abstract}



\section{Introduction}

Wireless networks (e.g cellular and Wi-Fi) are deployed under two modes of spectrum regulation: licensed (i.e. exclusive use) for cellular, and unlicensed spectrum (whereby sources have no interference protection by rule) as for Wi-Fi. This fundamental difference is reflected in the {\em access} mechanisms of the two regimes: scheduled time and/or frequency sharing by the cellular base-station, whereas Wi-Fi networks use a {\em distributed} random time-shared Distributed Coordination Function (DCF) access mechanism. The exploding growth of mobile network traffic (fed by high-end devices running bandwidth-hungry applications) has led network operators to consider various `offload' strategies in (typically indoor) hot-spots, whereby local traffic access is provided by broadband wireless LANs. Wi-Fi has been the network of choice, but with the emergence and maturation of LTE-Unlicensed technology, operators now have a choice of deploying one or both. 

A target for such coexistence operation is the 5 GHz UNII bands where a significant swath of additional unlicensed spectrum was earmarked by the FCC in 2014 \cite{FCC}. Two different specifications for unlicensed LTE operation have been proposed: LTE Licensed Assisted Access (LTE-LAA) and LTE Unlicensed (LTE-U). LTE-LAA (developed by 3GPP) integrates a Listen-Before-Talk (LBT) mechanism \cite{ETSILAA} similar to carrier sensing multiple access collision avoidance (CSMA/CA) for Wi-Fi, to enable spectrum sharing worldwide. LTE-U employs adaptive duty cycling - denoted as Carrier Sense Adaptive Transmission (CSAT) - to adapt the ON and OFF duration for LTE channel access \cite{LTEU}. LTE-U is proposed for regions where LBT is not required and is promoted by the LTE-U forum \cite{LTEU}. As currently specified, both LTE-LAA and LTE-U utilize carrier aggregation between a licensed carrier and an (additional) unlicensed carrier for enhanced data throughput on the downlink (DL), and all uplink traffic is transmitted on the licensed carrier. 

Our work is distinct from standardized LTE-U in several important aspects - we only consider unlicensed LTE with {\em fixed duty cycling} which we denote as LTE-DC\footnote{LTE-U implicitly assumes CSAT; thus LTE-DC is our shorthand for LTE-U with fixed duty cycling.}. Further, we {\em do NOT} consider the impact of {\em rate adaptation} on either Wi-Fi or LTE-U performance and assume saturation (full buffer) conditions. As such, we caution against any extrapolation or other untenable application of our throughput or fairness results to actual deployment scenarios that do not conform to our assumptions. 

\begin{figure*}[!htb]
\setlength{\belowcaptionskip}{-0.1in}
\begin{center}
\includegraphics[width=5.0in]{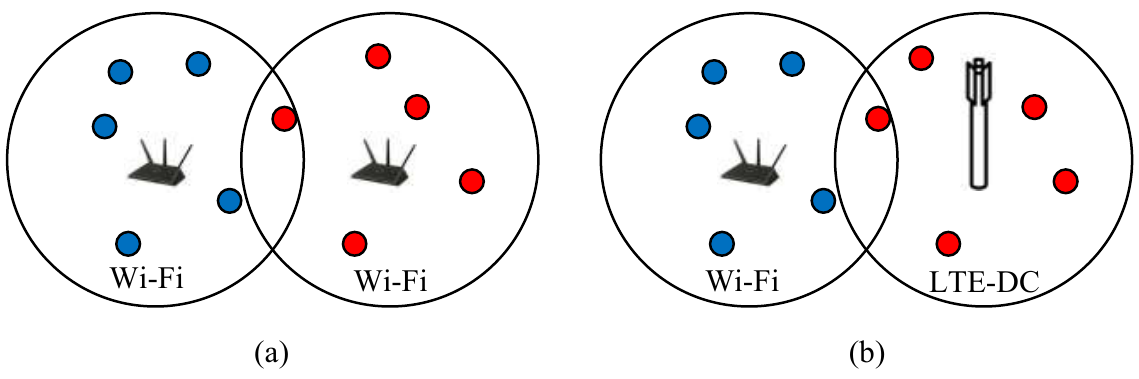}
 \caption{(a) coexistence of two Wi-Fi networks. (b) coexistence of a Wi-Fi network with an LTE-DC network which is transmitting in DL.}
 \label{fig: Diag}
\end{center}
\end{figure*}

Many industry inspired explorations regarding coexistence of Wi-Fi and LTE-U are based on simulations or experiments which are independently unverifiable. There is thus a need for a transparent and comprehensive analysis methodology for this problem, to which we attempt to make a significant contribution.
Our coexistence throughput model is backed by results from {\em actual network simulation} of the coexistence stack simulated within ns-3\footnote{The most popular open source network simulator for academic research, see www.nsnam.org.}, and paves the way for exploring Wi-Fi and LTE-DC fairness, a critical criterion for evaluation of any coexistence system.

The specific contributions of this work include: 
\begin{itemize}
\item  A new analytical model for throughput of Wi-Fi in coexistence with LTE-DC for full buffer load condition;
\item Simulating the coexistence scenario using the ns-3 simulator by developing the coexistence scenario and validating the analytical model based on the simulation results;
\item  Applying various structured definitions of fair sharing to this use-case and evaluating Wi-Fi access and throughput fairness for them.
\end{itemize}

This paper is organized as follows. Section II presents the related works by industry and academia. Section III describes the Wi-Fi and LTE-DC access mechanisms in coexistence network. In Section IV, the analytical throughput modeling of the coexistence network is presented. Section V illustrates the ns-3 simulation and numerical results for comparison. Section VI discusses the Wi-Fi fairness in coexistence with the LTE-DC. Section VII illustrates the numerical results of fairness investigation of the coexistence system. Finally, section VIII concludes the paper.


\section{Related Work}

An early work that explored 5 GHz LTE and Wi-Fi coexistence \cite{LTEWiFi_Mag} from a radio resource management perspective showed that Wi-Fi can be severely impacted by LTE transmissions in some conditions, suggesting the need for measures of fair coexistence. Thereafter Nokia Research \cite{LTE_Nokia} proposed a  bandwidth sharing mechanism whereby the impact on Wi-Fi network throughput can be controlled by restricting LTE activity. In situations where a large number of Wi-Fi users try to access the network, users may spend a long time in back-off (i.e. medium is idle); if LTE could exploit these silent times, overall bandwidth utilization efficiency could increase without negatively impacting Wi-Fi performance. In \cite{LTEWiFi_Per}, a performance evaluation of LTE and Wi-Fi coexistence was conducted via simulation which showed that while LTE system performance is slightly affected, Wi-Fi is significantly impacted by LTE transmissions. Wi-Fi channel access is most often blocked by LTE transmissions, causing the Wi-Fi nodes to stay in the listen mode more than 96\% of the time. 

In \cite{Google_LTEU}, Google's investigation shows that in many circumstances, LTE-U coexists poorly with Wi-Fi in the 5 GHz band. The underlying causes include:  a) LTE-U's duty-cycling causes LTE transmissions to begin abruptly, often in the middle of Wi-Fi transmissions, interrupting them and causing Wi-Fi to ratchet down the transmission rates via rate control in response to increased error rates; and b) the lack of an effective coexistence mechanism in scenarios where LTE-U and Wi-Fi devices hear each other at moderate but non-negligible power levels. In \cite{Cablelabs}, the research noted that LTE-U and Wi-Fi coexistence is a balancing act between throughput and latency. Either throughput or latency of a co-channel Wi-Fi network is negatively affected if the LTE  duty cycle period is too low or too high, respectively. They also illustrate that with 50\% duty cycling, the Wi-Fi throughput would be affected by more than half. On the other hand, Qualcomm \cite{Qualcomm} investigated the coexistence of Wi-Fi with LTE-U through simulation and showed that significant {\em throughput gain} can be achieved by aggregating LTE across licensed and unlicensed spectrum; further (and importantly), this throughput improvement does not come at the expense of degraded Wi-Fi performance and both technologies can fairly share the unlicensed spectrum. 

In \cite{Broadcom_LTEU} report to FCC, Wi-Fi coexistence with LTE-U was measured by impact on Wi-Fi throughput, latency, and VoIP dropped calls via simulation. The study concluded that LTE-U (as prescribed) does not meet the `fair' coexistence criterion and additional measures are necessary. In \cite{Huawi}, Huawei discussed the fundamental differences in the physical and Medium Access Control (MAC) layer design between LTE and Wi-Fi, that may negatively impact the channel occupancy of co-channel Wi-Fi, especially in some high-load cases. However, LTE Pico performs much more robustly even with high-load interfering access point (AP) nearby. Several factors contribute to this result, including link adaptation and Hybrid Automatic Repeat Request (HARQ) retransmission in LTE.

In \cite{LTEU_fairLeith1}, the fairness of Wi-Fi airtime in the coexistence of Wi-Fi/LTE-LAA LBT and Wi-Fi/LTE-U with CSAT is investigated. The paper shows that both mechanisms can provide the same level of fairness to Wi-Fi transmissions if a suitable proportional fair rate allocation is used. Consequently, the choice between using CSAT and LBT is a decision driven by the LTE operators. In \cite{AssocFair}, the effect of large LTE-U duty cycle on the association process of Wi-Fi was explored. A National Instrument (NI) experimental set-up was used to show that a significant percentage of Wi-Fi beacons will either not be transmitted in a timely fashion or will not be received at the LTE-U BS which makes it difficult for the LTE-U BS to adapt its duty cycle in response to the Wi-Fi usage. The results in this paper illustrated that in order to maintain association fairness, the LTE-U should not transmit at the maximum duty cycles of 80\% even if it deems the channel to be vacant.

In summary, industry-driven research shows a) are taken negative consequences occur for Wi-Fi with the proposed LTE-DC coexistence mechanisms, while others b) claim that the fair coexistence is feasible with necessary tweaks or enhancements. 
Transparent analytical investigations on this topic have been sparse. One notable effort is \cite{Babaei_LTEU} where channel access probability for Wi-Fi stations is computed in presence of LTE-DC. While they model the (expected) decrease in Wi-Fi channel access due to LTE-DC, the additional collision probability of Wi-Fi caused by interference with the onset of an LTE-U ON period is not considered. The analytical model in \cite{Amr_LTEU} for the coexistence of Wi-Fi and LTE-DC does not provide any close form expressions (for collision probability and throughput) and their scenario does not conform to the real specification of LTE-DC \cite{LTEU}. We hope that our contribution to this important problem validated by actual network simulation provides a basis for resolving some of the importance coexistence issues.

\begin{table}
\caption{Glossary.}
\label{table: GlosPara}
\begin{center}
\begin{tabular}{|c|c|}
\hline
Parameter  & Definition \\
\hline
\hline
 $W_0$ & Wi-Fi minimum contention window \\
\hline
 $m$ & Wi-Fi maximum retransmission stage \\
\hline
 PhyH & preamble with physical header  \\
\hline
 MACH & MAC header \\
\hline
 ACK & Acknowledgment length \\
\hline
 $\sigma$ & Wi-Fi slot time \\
 \hline
 $\delta$ & propagation delay \\
\hline
 DIFS & distributed interframe space \\
\hline
 SIFS & short interframe space \\
 \hline
 $T_{on}$ & LTE-DC ON period \\
\hline
 $T_{off}$ & LTE-DC OFF period \\
 \hline
 $T_{C}$ & LTE-DC total cycle period \\
\hline
 $\alpha$ & LTE-DC duty cycle \\
\hline
 $N_B$ & Wi-Fi packet data portion size \\
\hline
 $T_d$ & Wi-Fi data portion duration \\
\hline
 CCA & Clear Channel Assessment \\ 
\hline
 $n_w$ & Number of Wi-Fi APs in coexistence \\
 \hline
 $r_0$ & MCS0 data rate \\
\hline
 $r_w$ & Wi-Fi data rate \\
\hline
 $r_l$ & LTE-DC data rate \\
\hline
 $P_{c,w}$ & Wi-Fi Collision Probability \\
\hline
 $\tau_w$ & a Wi-Fi station transmission probability per slot \\
\hline
 $P_{c,w-l}$ & Collision probability from LTE-DC to Wi-Fi \\
\hline
 $P_{c,t}$ & total collision probability \\
\hline
 $n_k$ & maximum number of packets fit in one OFF period \\
\hline
 $L_b(k)$ & lower bound of the interval for contention \\
\hline
 $U_b(k)$ & upper bound of the interval for contention \\
\hline
 $p'_h(k)$ & probability of $k$-th packet hitting the ON edge \\
\hline
 $P'_s(k)$ & successful transmission probability of $k$-th packet \\
\hline
 $E_n$ & average number of packets fit in the OFF period\\
\hline
 $P_{trw}$ & Wi-Fi transmission probability \\
\hline
 $P_{sw}$ &  Wi-Fi successful transmission probability \\
\hline
 $Tput_{w}$ & Wi-Fi throughput \\
\hline
 $Tput_{l}$ & LTE-DC throughput \\
\hline
 $\tau_{wo}$ & a Wi-Fi station transmission probability per slot\\ & in Wi-Fi only network \\
\hline
 $P_{c,wo}$ & Collision probability of Wi-Fi in Wi-Fi only network \\
\hline
 $Tput_{wo}$ & Wi-Fi throughput in Wi-Fi only network \\
\hline
 $T_{sw}$ & expected duration of a successful packet \\
\hline
 $T_{cw}$ & expected duration of a collided packet \\
\hline
\end{tabular}
\end{center}
\vspace{-0.2in}
\end{table}

\begin{figure*}[!htb]
\setlength{\belowcaptionskip}{-0.1in}
\begin{center}
\includegraphics[width=4.3in]{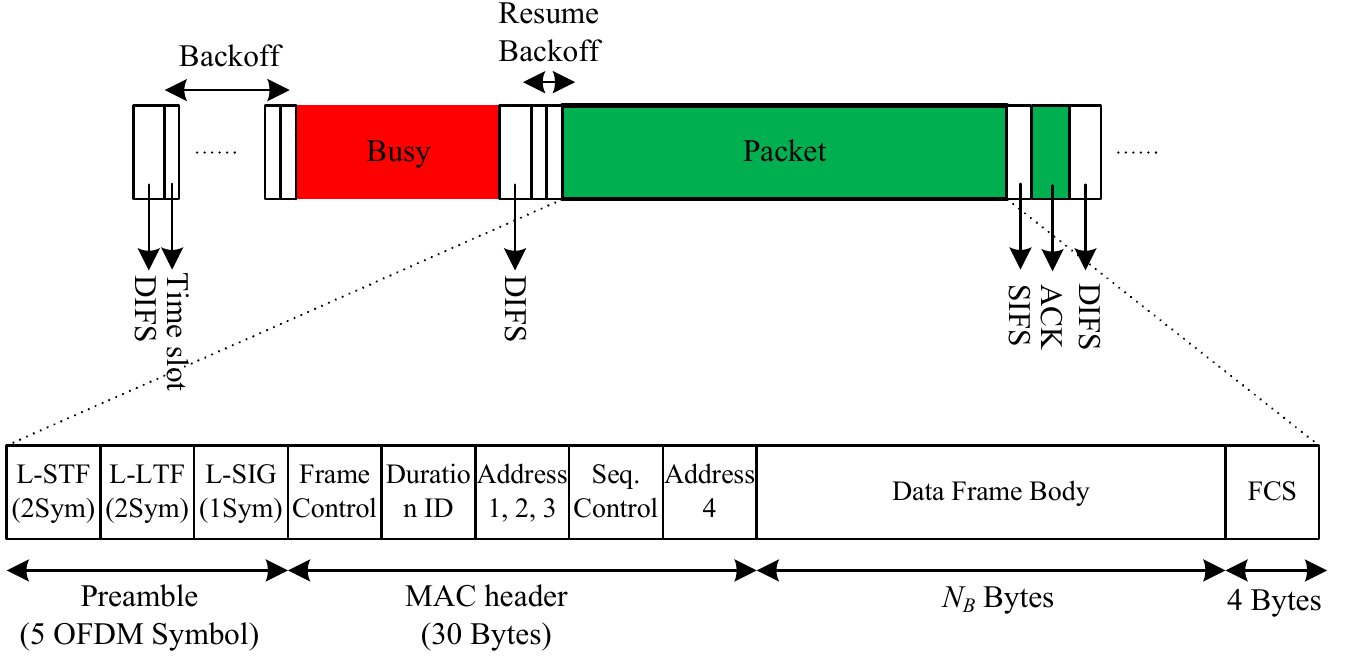}
 \caption{Wi-Fi CSMA/CA contention and frame transmission. The Wi-Fi frame structure with Preamble, MAC header, and data portion.}
 \label{fig: WiFitime}
\end{center}
\end{figure*}


\section{Coexistence of LTE-DC and Wi-Fi: MAC Protocol Mechanisms}

A brief review of the  Wi-Fi DCF and LTE-DC MAC is presented, to highlight their basic differences - LTE-DC is a scheduled time division multiple access (TDMA) based while Wi-Fi is random access CSMA/CA.

\subsection{Wi-Fi DCF}
\label{sec: WiFiDCF}

\begin{figure}[!htb]
\setlength{\belowcaptionskip}{-0.1in}
\begin{center}
\includegraphics[width=2.7in]{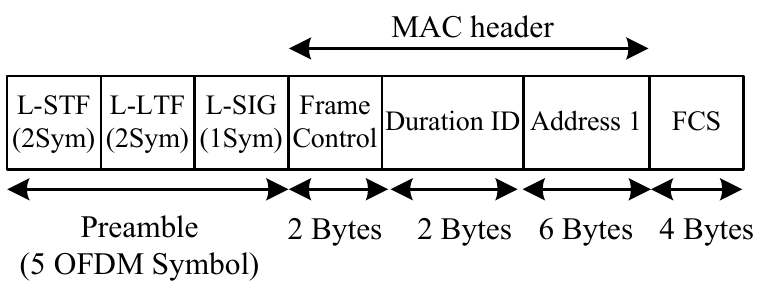}
 \caption{Wi-Fi ACK frame structure}
 \label{fig: ACKframe}
\end{center}
\end{figure}

The Wi-Fi distributed coordination function (DCF) MAC employs CSMA/CA \cite{std80211} as illustrated in Fig.~\ref{fig: WiFitime}. Each node attempting transmission must first ensure that the medium has been idle for a duration of DCF Interframe Spacing (DIFS) using the energy detection (ED)\footnote{The ability of Wi-Fi to detect any external (out-of-network) signal using an energy detector.} and carrier sensing (CS)\footnote{The ability of Wi-Fi to detect and decode an incoming Wi-Fi signal preamble.} mechanism. When either of ED or CS is true, the clear channel assessment (CCA) flag is indicated as busy. If the channel has been detected idle for DIFS duration and the station is not accessing immediately after a successful transmission, it transmits. Otherwise, if the channel is sensed busy (either immediately or during the DIFS) or the station is again seeking channel access after a successful transmission, the station persists with monitoring the channel until it is measured idle for a DIFS, then selects a random back-off duration (in units of slot time  $\sigma=9\mu s$ ) and counts down. Specifically, a station selects a back-off counter uniformly at random in the range of $[0, 2^i W_0 - 1]$ where the value of $i$ (the back-off stage) is initialized to 0 and $W_0$ is the {\em minimum contention window} chosen initially.  Each failed transmission due to packet collision\footnote{A collision event occurs if and only if two nodes select the same back-off counter value at the end of a DIFS period (if there is no hidden terminal).} results in incrementing the back-off stage by $1$ (binary exponential back-off or BEB) and the node counts down from the selected back-off value. During back-off, a node decrements the counter every slot duration as long as no other transmissions are detected. If during countdown a transmission is detected, the counting is paused; nodes continue to monitor the busy channel until it goes idle for DIFS period before the back-off countdown is resumed. Once the counter hits zero, the node transmits a packet. Any node that did not complete its countdown to zero in the current round, carries over the back-off value and resumes countdown in the next round.  Once a transmission has been completed successfully, the value of $i$ is reset to 0. The maximum value of back-off stage $i$ is $m$; it stays in $m$-th stage for one more unsuccessful transmission (for a total of $2$ attempts at back-off stage $m$). If the final transmission is unsuccessful, the node drops the packet and resets the back-off stage to $i=0$. For any successful transmission, the intended receiver will transmit an acknowledgment frame (ACK) after a Short Interframe Spacing (SIFS) duration post reception; the ACK frame structure is shown in Fig.~\ref{fig: ACKframe} which consists of preamble and MAC header. The ACK frame chooses the highest basic data rate (6 Mbps, 12 Mbps, or 24 Mbps) for transmitting the MAC header which is smaller than the data rate used for data transmission. 

\subsection{LTE-DC}
\label{sec: LTEU}

\begin{figure*}[!htb]
\setlength{\belowcaptionskip}{-0.1in}
\begin{center}
\includegraphics[width=4.6in]{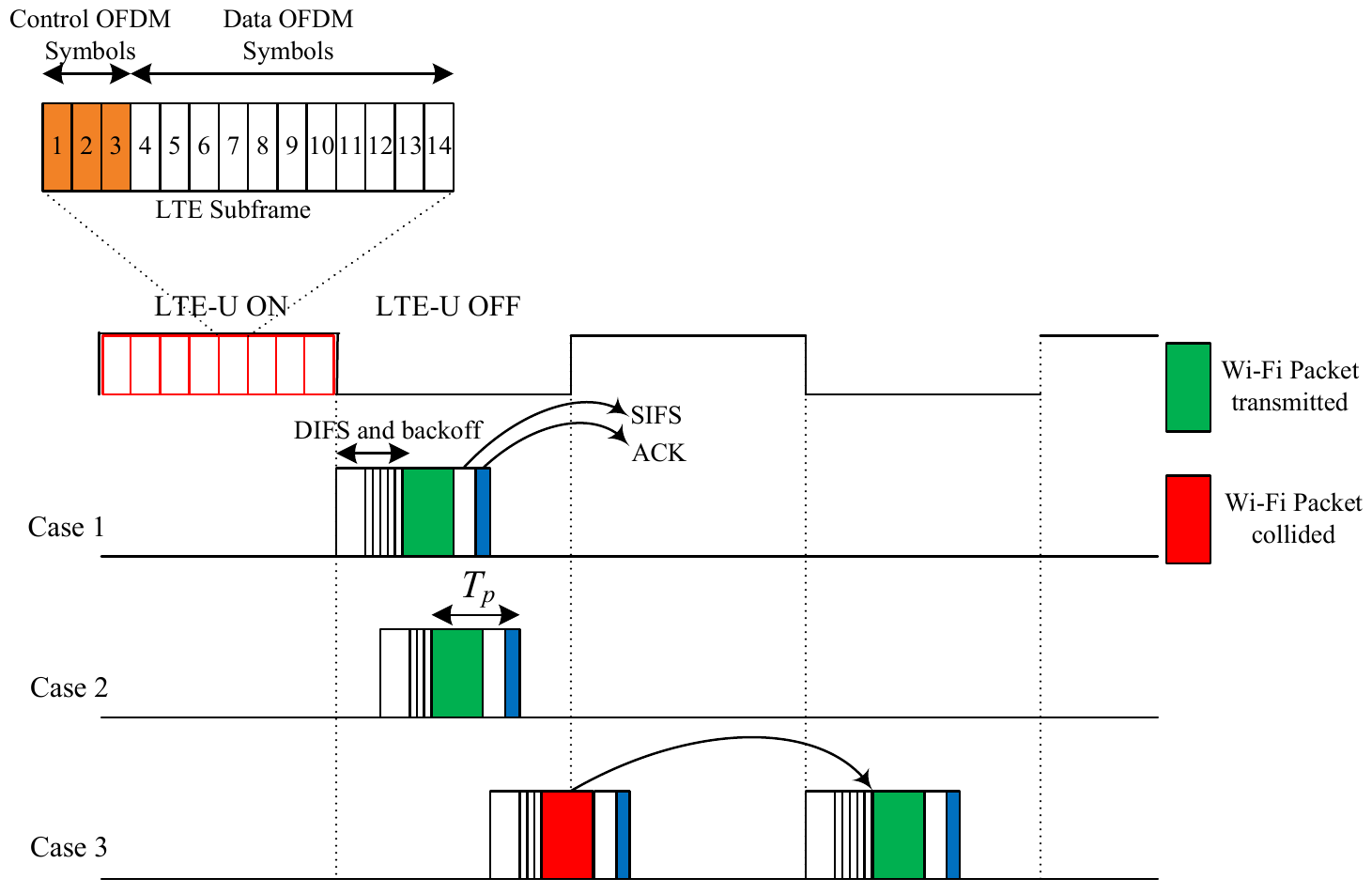}
 \caption{LTE-DC coexisting with Wi-Fi which causes the overlap of Wi-Fi frame with LTE-DC ON edge}
 \label{fig: coex}
\end{center}
\end{figure*}

LTE-DC uses a duty-cycling approach (i.e. alternating the ON and OFF period, where the LTE evolved Node B (eNB) is allowed to transmit only during the ON duration) where the duty cycle (ratio of ON duration to one cycle period) is determined by perceived Wi-Fi usage at the LTE-DC eNB, using carrier sensing. During the ON period, the LTE-U eNB schedules DL transmissions to UEs, unlike Wi-Fi in which transmissions are governed by the CSMA/CA process. Fig.~\ref{fig: coex} shows the LTE-DC transmission for the duty cycle of 0.5. If a single Wi-Fi AP is detected, the maximum duty cycle could be set at 0.5. The LTE-U Forum specifications \cite{LTEU} provide limits on the a) actual durations of ON state (both minimum and maximum) and b) the minimum OFF duration. Specifically, the maximum ON duration is 20 ms and the minimum ON duration is 4 ms as long as there is data in LTE-DC eNB's buffer. The minimum OFF duration of the eNB is 1 ms. Therefore, if no active Wi-Fi stations are detected, an LTE-DC eNB can start transmissions with the maximum ON period of 20 ms and OFF period of 1 ms, resulting in a maximum duty cycle of 95\%. 

LTE-DC uses the base LTE subframe structure, i.e., the subframe length of 1 ms; each sub-frame consists of two $0.5$ ms slots. Each subframe consists of 14 OFDM symbols as indicated in Fig.~\ref{fig: coex}, of which 1 to 3 are Physical Downlink Control Channel (PDCCH) symbols and the rest are Physical Downlink Shared Channel (PDSCH) data. LTE-DC eNBs start downlink transmissions synchronized with slot boundaries, for (at least) one subframe (2 LTE slots) duration. After transmission, the intended receiver (or receivers) transmits the ACK on uplink via the {\em licensed band} if the decoding is successful.  

In LTE, a Resource Block (RB) is the smallest unit of radio resource which can to to a UE, equal to 180 kHz bandwidth over a Transmission Time Interval (TTI) of one subframe (1 ms). Each RB of 180 kHz bandwidth contains 12 sub-carriers, each with 14 OFDM symbols, equaling 168 Resource Elements (REs). Depending upon the modulation and coding schemes (QPSK, 16-QAM, 64-QAM), each symbol or resource element in the RB carries 2, 4 or 6 bits per symbol, respectively. In the LTE system with 20 MHz bandwidth, there are 100 RBs available.


\section{Analytical Throughput Modeling of the Coexistence Network}

The fundamental contribution to a new coexistence analysis (i.e. prediction of Wi-Fi throughput in the presence of LTE-DC eNB) is based on the Two-Dimensional (2-D) Markov model. The key changes come via the amended Wi-Fi packet collision probabilities due to the presence of LTE-DC eNB transmissions; this determines the effective available airtime for Wi-Fi transmission and the way throughput is calculated during LTE-DC OFF period. We assume that both Wi-Fi and LTE-DC systems transmit on a 20 MHz channel; there are $n_w$ Wi-Fi nodes (AP and stations) which are transmitting both in down and uplinks with full buffer (i.e. saturation) and one LTE-DC eNB which is transmitting only in DL to LTE stations. 

\begin{figure}[t]
\setlength{\belowcaptionskip}{-0.1in}
\begin{center}
\includegraphics[width=3.6in]{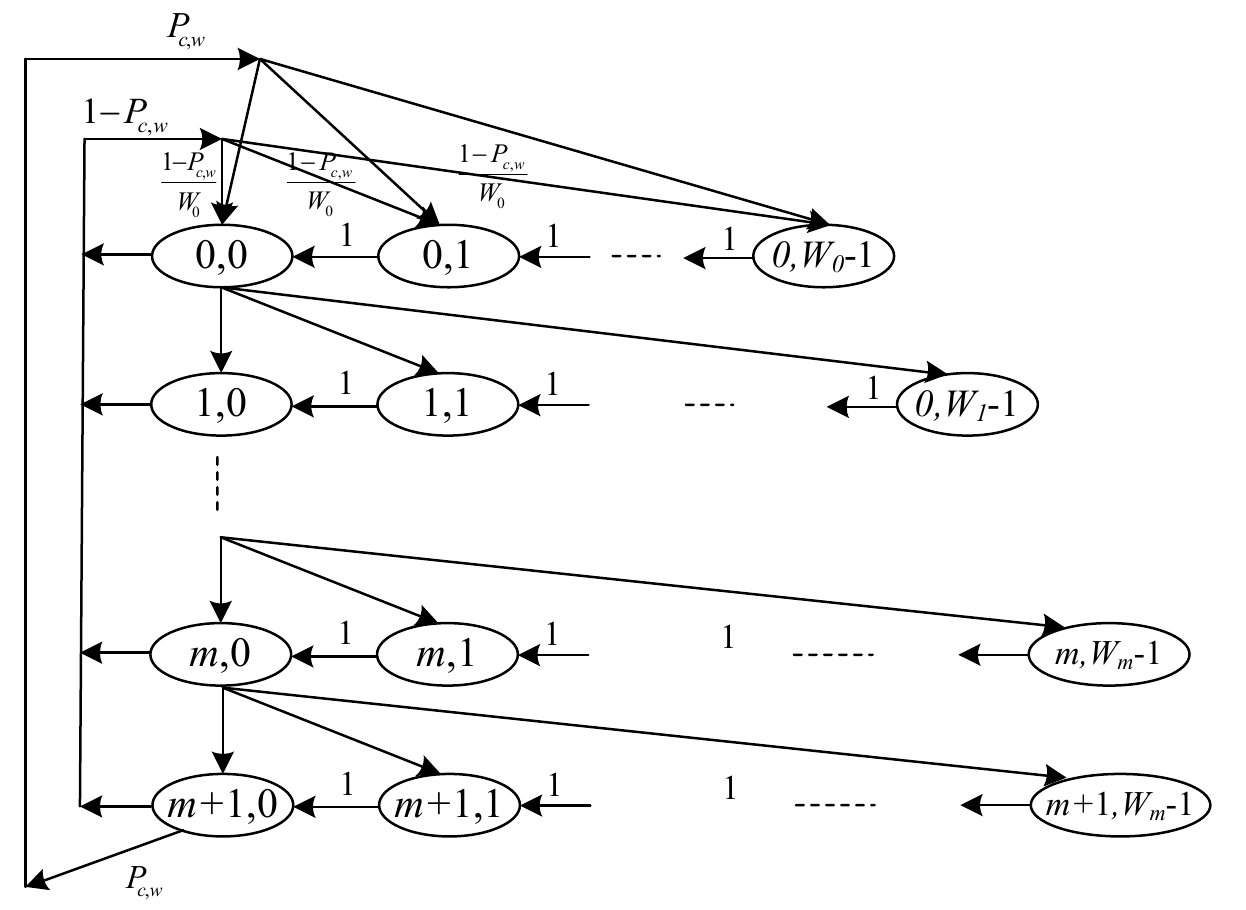}
 \caption{Markov chain model for the Wi-Fi DCF with binary exponential back-off}
 \label{fig: MarkWi-Fi}
\end{center}
\end{figure}

The 2-D Markov chain model for Wi-Fi DCF is shown in Fig.~\ref{fig: MarkWi-Fi} for the saturated nodes. Let $\{s(t)=j,b(t)=k\}$ denote the possible states in the Markov chain, where $s(t)$ is the retransmission stage and $b(t)$ the back-off counter value. In \cite{Bianchi}, when the back-off stage reaches the maximum value (i.e. $m$), it stays in $m$ forever. However, in Wi-Fi when the maximum value is reached, the back-off stage stays at $m$ for one more attempt, i.e. $m+1$; then it resets to zero in case of an unsuccessful transmission. The Markov chain and corresponding one step transition probabilities in \cite{Bianchi} are modified based on Section \ref{sec: WiFiDCF} as follows. Considering the stationary distribution for the Markov model as $b_{j,k}=\lim_{t\to \infty}P\{s(t)=j,b(t)=k\}, j \in (0,m+1), k \in (0,W_i-1)$, the modified one step transition probability of the Markov chain is,
\begin{align}
 \left \{ \hspace{-0.2in} \begin{array}{c c c}
&P\{j,k|j,k+1\} = 1, ~~~ k \in (0,W_i-2) ~~~ j \in (0,m+1) \\
&P\{0,k|j,0\} = \frac{1-P_{c,w}}{W_0}, ~~~ k \in (0,W_0-1) ~~~ j \in (0,m+1) \\
&P\{j,k|j-1,0\} = \frac{P_{c,w}}{W_i}, ~~ k \in (0,W_i-1) ~~~ j \in (1,m+1) \\
&P\{0,k|m+1,0\} = \frac{P_{c,w}}{W_0}, ~~~ k \in (0,W_0-1) ~~~~~~~~~~~~~~~~ 
\end{array} \right.
\label{eq: trans1}
\end{align}
where $P_{c,w}$ is the collision probability of Wi-Fi nodes, $W_0$ is the minimum contention window size in CSMA/CA, $W_i=2^iW_0$ is the contention window size at the retransmission stage $i$, and $i=m$ is the maximum retransmission stage (i.e., $i=j$ for $j \le m$ and $i=m$ for $j > m$).

To simplify the calculation, we introduce the following variables derived from (\ref{eq: trans1}) and using the Markov Chain properties: 
\begin{align}
 \left \{ \begin{array}{c c}
&b_{j,0}=P_{c,w} b_{j-1,0}, ~~~~~ 0 < j \le m+1\\
&b_{j,0}=P_{c,w}^j b_{0,0}, ~~~~~~~~ 0 \le j \le m+1\\
&b_{0,0} = P_{c,w} b_{m+1,0}+(1-P_{c,w}) \sum_{j=0}^{j=m+1} b_{j,0}
\label{eq: stat1w}
\end{array} \right.,
\end{align}
the last equation implies that,
\begin{equation}
\sum_{j=0}^{j=m+1} b_{j,0}=\left(\frac{1-P_{c,w}^{m+2}}{1-P_{c,w}}\right ) b_{0,0}.
\label{eq: stat2w}
\end{equation} 

In each retransmission stage, the back-off transition probability is 
\begin{equation}
b_{j,k}=\frac{W_i-k}{W_i}b_{j,0}, ~~ 0 \le j \le m+1, ~ 0 \le k \le W_i-1.
\label{eq: stat3w}
\end{equation}

We can derive $b_{0,0}$ by the normalization condition, i.e., 
\begin{equation}
\begin{split}
&\sum_{j=0}^{m+1}\sum_{k=0}^{W_i-1}b_{j,k}=1,\\
&b_{0,0}=\frac{2}{W_0\left(\frac{(1-(2P_{c,w})^{m+1})}{(1-2P_{c,w})}+\frac{2^{m}\left(P_{c,w}^{m+1}-P_{c,w}^{m+2}\right)}{(1-P_{c,w})}\right) +\frac{1-P_{c,w}^{m+2}}{1-P_{c,w}}}.
\label{eq: normw}
\end{split}
\end{equation}

Hence, the probability that a node transmits in a time slot is calculated using (\ref{eq: stat2w}) and (\ref{eq: normw}) as,
\begin{equation}
\begin{split}
&\tau_w = \sum_{j=0}^{m+1}b_{j,0}=\\&\frac{2}{W_0\left(\frac{(1-(2P_{c,w})^{m+1})(1-P_{c,w})+2^{m}\left(P_{c,w}^{m+1}-P_{c,w}^{m+2}\right)(1-2P_{c,w})}{(1-2P_{c,w})(1-P_{c,w}^{m+2})}\right) +1}.
\label{eq: tauw}
\end{split}
\end{equation}

The collision probability of a Wi-Fi station with at least one of the other remaining ($n_w-1$ Wi-Fi) stations is given by
\begin{equation}
P_{c,w} =1-(1-\tau_w)^{n_w-1},
\label{eq: Pw}
\end{equation}
where $P_{c,w}$ is coupled to $\tau_w$. 

\begin{figure}[t]
\setlength{\belowcaptionskip}{-0.1in}
\begin{center}
\includegraphics[width=3.6in]{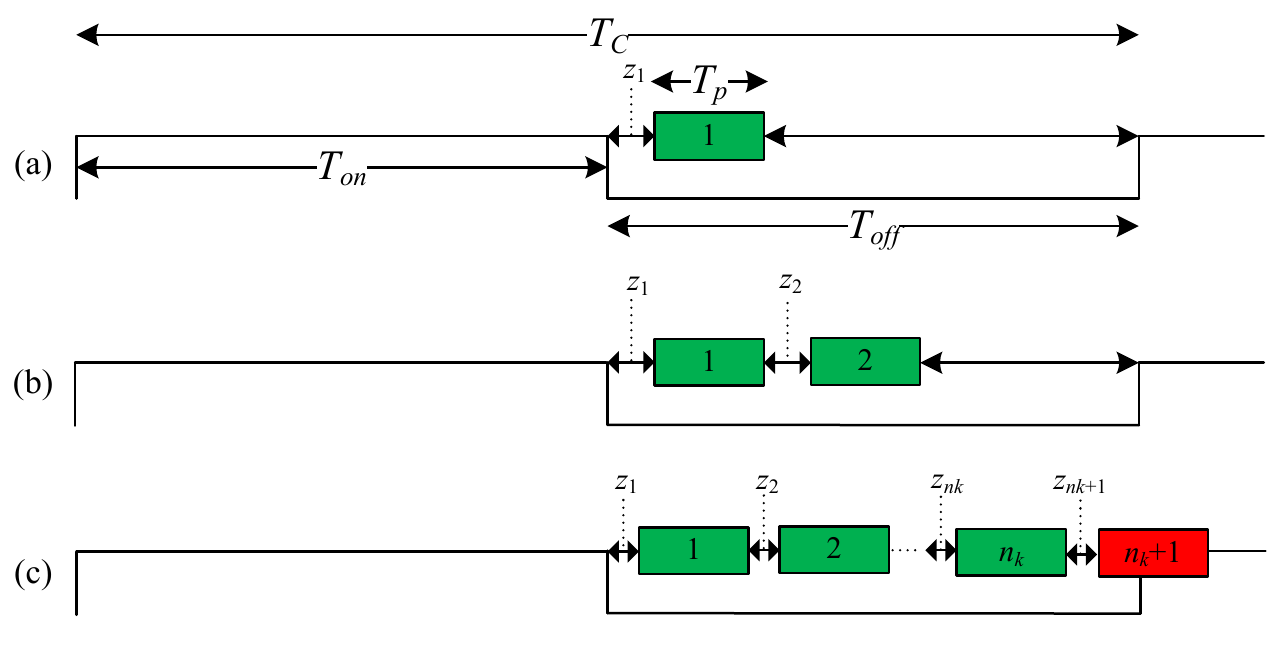}
 \caption{Illustration: Modeling collision and successful packet transmission at the end of LTE-DC OFF period. $z_i$ is the random backoff number and the backoff time is backoff number multiplied by the time slot $\sigma z_i$.}
 \label{fig: Pc}
\end{center}
\end{figure}

\subsection*{Modeling Collision between Wi-Fi \& LTE-DC}

Inter-network collisions between Wi-Fi packets and LTE-DC occurs at the transition from OFF to ON period as illustrated in Case 3 of Fig.~\ref{fig: coex}. The occurrence of this event depends on various factors: the total airtime for a Wi-Fi packet transmission equaling the sum of the Wi-Fi data packet duration, SIFS duration, and ACK duration, denoted as $T_p=\text{MACH}+\text{PhyH}+T_d+\text{SIFS}+\text{ACK}$. We assume that any overlap of a Wi-Fi packet airtime $T_p$ duration with the LTE-DC ON period causes the Wi-Fi packet collision. For calculating the collision probability from LTE-DC to Wi-Fi, we initially assume that {\em one} Wi-Fi station coexisting with LTE-DC (hence no contention among the Wi-Fi network), and thereafter extend it to multiple Wi-Fi stations (to capture the Wi-Fi contention). 

The probability of a collision between a Wi-Fi station transmission and LTE-DC downlink is calculated as
\begin{equation}
P_{c,w-l} =\sum_{k=1}^{n_k+1}P(C|H=k)p'(H=k)=\sum_{k=1}^{n_k+1}\frac{1}{k}p'_h(k),
\label{eq: Plte}
\end{equation}
where $n_k=\lfloor \frac{T_{off}}{T_p} \rfloor$ is the maximum number of consecutive Wi-Fi packets that fit within an OFF duration. $H$ is the random variable for the overlap probability of the Wi-Fi packets transmitted in OFF period with LTE-DC ON edge and $C$ is the collision of the last transmitted packet during OFF period. Thus, $P(C|H=k)=\frac{1}{k}$ is the collision probability reflecting the fact that only the last transmitted packet (among $k$ transmitted packets during the OFF period) is lost due to collision. To compute $p'(H=k)=p'_h(k)$, that represents the probability that the $k$-th packet overlaps with the LTE-DC ON edge, refer to Fig.~\ref{fig: Pc} for an illustrative example. Suppose that $T_{off}=5$ ms and packet airtime $T_p=2.3$ ms. In this case, the transmitting Wi-Fi station either starts from $[0,W_0-1]$ contention window or $[0,2W_0-1]$ in case of collision with the previous LTE-DC ON edge. To simplify matters, we assume that the first packet in an OFF period chooses the random backoff from $[0,2W_0-1]$ because the last packet in the previous OFF period is lost (with high probability). This assumption simplifies the derivation and we will show in the following sections that this approximation is valid for most cases by comparing the collision probability from this theoretical model and ns-3 simulation results. Thus, the probability that the first transmitted packet in OFF period overlaps the next ON edge is 
\begin{equation}
\begin{split}
p'_{h}(1)=P(L_b(1)<z_1<U_b(1)),
\end{split}
\label{eq: Pc1}
\end{equation}
where $z_1$ represents the random back-off value (in terms of number of contention time slots) that will lead to a packet transmission collision.  $\sigma$ is the time slot duration, $L_b(1)=\lfloor (T_{off}-T_P-\text{DIFS})/\sigma \rfloor$ and $U_b(1)=\lfloor (T_{off}-\text{DIFS})/\sigma \rfloor$ are the upper and lower bounds of the collision interval duration. The random back-off number of the first packet follows Uniform distribution, i.e. $z_1 \; \sim \; U [0,2W_0-1]$. Thus, the probability that the first packet overlaps with the ON edge is 
\begin{equation}
\begin{split}
p'_{h}(1)=
\begin{cases}
0,~ \text{for}~ L_b(1)>2W_0-1 \\
\sum_{z_1=L_b(1)}^{2W_0-1}\left(\frac{1}{2W_0}\right),~ \text{for}~0\le L_b(1)\le 2W_0-1\\
1,~ \text{for}~ L_b(1)<0
\end{cases}
\label{eq: Pc2}
\end{split}
\end{equation}

For the second packet onward in each OFF period, the station chooses the random backoff uniformly from $[0,W_0-1]$ because there is no collision, i.e. first retransmission stage of Markov model. Assuming $k-1$ packets are successfully transmitted, the probability of collision for $k$-th packets is calculated as:
\begin{equation}
\begin{split}
&p'_{h}(k)=\sum_{z_1=0}^{\min\{2W_0-1,U_b(k)\}}p(z_1)\times\\ &P(\max\{0,L_b(k)-z_1\}<Z(k)<\min\{W_s,U_b(k)-z_1\}|z_1),
\end{split}
\label{eq: Pc3}
\end{equation}
where $L_b(k)=\lfloor (T_{off}-k(T_P+\text{DIFS}))/\sigma \rfloor$ and $U_b(k)=\lfloor (T_{off}-(k-1)T_P-k\text{DIFS})/\sigma \rfloor$ are the lower and upper bounds of interval for $k>1$, $W_s=(k-1)W_0-1$ is the maximum sum of backoff numbers, and $Z(k)=\sum_{i=2}^{k}z_i$ is the sum of the backoff values from the second to the $k$-th packet. Since the $z_i$ are mutually Independent and Identically Distributed (IID), the Probability Density Function (PDF) of $Z(k)$ is the $k-1$-fold convolution of uniform PDF $p(z_k)=\frac{1}{W_0}$. 

Thus, $P_{c,w-l}$ which is the collision probability from LTE-DC to Wi-Fi depends on $T_{off}$ and airtime of the packet, that means collision probability is a function of $\alpha=\frac{T_{on}}{T_{C}}$ - the duty cycle of LTE-DC ON period - and $T_{C}$. The collision caused by LTE-DC ($P_{c,w-l}$) and collision from other Wi-Fi stations ($P_{c,w}$) is next combined to calculate the total probability of collision as:
\begin{equation}
\begin{split}
P_{c,t}&=1-(1-P_{c,w})(1-P_{c,w-l})\\
&=1-(1-\tau_w)^{n_w-1}(1-P_{c,w-l}).
\label{eq: Pwp}
\end{split}
\end{equation}

In order to compute the $P_{c,t}$ and $\tau_w$ for more than one stations ($n_w>1$) in the network ($\tau_w$ depends on $P_{c,t}$, so it is a function of $\alpha$ and $T_{C}$ as well), we first jointly solve eq. (\ref{eq: tauw}) and (\ref{eq: Pwp}), where for the coexistence of Wi-Fi and LTE-U, $P_{c,t}$  is replaced by $P_{c,w}$ in eq. (\ref{eq: tauw}). 

The transmission probability of Wi-Fi is the probability that at least one of the $n_w$ stations transmit a packet during a time slot:
\begin{equation}
P_{trw} =1-(1-\tau_w)^{n_w}.
\label{eq: Ptrw}
\end{equation}

The successful transmission probability of Wi-Fi is the event that exactly one of the $n_w$ stations makes a transmission attempt given that at least one of the Wi-Fi APs transmit:
\begin{equation}
P_{sw} =\frac{n_w \tau_w (1-\tau_w)^{n_w-1}}{P_{trw}}.
\label{eq: Psw}
\end{equation}

The average throughput of Wi-Fi in co-existence is calculated as:
\begin{equation}
\begin{split}
&Tput_{w} = \frac{ E_n T_d P_{sw}}{T_{C}}~r_w,
\label{eq: tputwc}
\end{split}
\end{equation}
where $r_w$ is the Wi-Fi data rate. $E_n$ is the average number of successfully transmitted packets in the OFF period, and is multiplied by $T_d$ to calculate the effective duration of data transmission in one cycle ($T_{C}$). $P_{sw}$ in the numerator captures the effect of contention/collision between Wi-Fi nodes and collision between Wi-Fi packet and LTE-DC ON edge. The $E_n$ is calculated as:
\begin{equation}
E_n=\sum_{k=1}^{k=n_k}k(P'_s(k)-P'_s(k+1)),
\label{eq: En}
\end{equation}
where $P'_s(k)$ is the transmission probability that a sequence of $k$ Wi-Fi packets are successfully transmitted in one cycle in coexistence with LTE-DC and $(P'_s(k)-P'_s(k+1))$ is the probability that exactly the $k$-th packet is successful.

Thus for a {\bf single station} in the Wi-Fi network, the successful transmission probability of the {\em first} packet is calculated as:
\begin{equation}
\begin{split}
&P'_{s}(1)=P(0<z_1<\min\{L_b(1),2W_0-1\})\\
&=\begin{cases}
1,~ \text{for}~ L_b(1)>2W_0-1 \\
\sum_{z_1=0}^{min\{L_b(1),2W_0-1\}}\frac{1}{2W_0},~ \text{for}~ 0 \le L_b(1)\le 2W_0-1\\
0,~ \text{for}~ L_b(1)< 0
\end{cases}
\label{eq: Ps1}
\end{split}
\end{equation}
and the successful probability of transmission for all ($k>1$) subsequent packets is calculated as:
\begin{equation}
\begin{split}
P'_{s}(k)&=\sum_{z_1=0}^{\min\{2W_0-1,U_b(k)\}}p(z_1)\times\\ &P(0<Z(k)<\min\{W_s,L_b(k)-z_1\}|z_1).
\end{split}
\label{eq: Ps2}
\end{equation}

For {\bf more than one Wi-Fi station} in the network ($n_w>1$), the $P'_s(k)$ should be re-computed; the probability that at least one node transmits - $P_{trw}$ - must now capture the effect of collision with other Wi-Fi stations and the collision with LTE-DC downlink. In this case, the random number of backoff time slots $z_k$ for $k$-th packet as shown in Fig. 6 follows the Geometric distribution, i.e. $P (z_k = i)=P_{trw}(1-P_{trw})^i,~i=0,1,..., \infty$. Hence, the total contention time slots in any generic OFF period  $Z'(k)=\sum_{j=1}^{k} z_j$ is the sum of $k$ IID Geometric random variables, resulting in a negative Binomial distribution:
\begin{equation}
P (Z'(k) = i)=\binom{i+k-1}{k-1}P_{trw}^k(1-P_{trw})^i,
\label{eq: Ps3}
\end{equation}
where $i$ is the total number of contention time slots and $k$ is the number of transmitted packets, in the OFF period. So the success probability for $k$ packets is calculated as:
\begin{equation}
\begin{split}
P'_{s}(k)&=P(Z'(k) \le L_b(k)-k) \\
&=\sum_{i=0}^{L_b(k)-k}\binom{i+k-1}{k-1}P_{trw}^k(1-P_{trw})^i.
\label{eq: Ps4}
\end{split}
\end{equation}

The throughput of the LTE-DC considering the LTE-DC ON time, duty cycle, and proportion of data symbols in each subframe is calculated as,
\begin{equation}
Tput_{l}(\alpha) = \frac{\frac{13}{14} T_{on}}{ T_{C} }~r_l=\frac{13}{14}\alpha r_l,
\label{eq: tputl}
\end{equation}
where $\frac{13}{14} T_{on}$ is the fraction of the LTE-DC TXOP in which the data is transmitted, i.e. $1$ PDCCH symbol in a subframe with 14 OFDM symbols is considered, and $r_l$ is the LTE-LAA data rate. The effect of Wi-Fi packet collision to LTE-DC is very minor as the collision from Wi-Fi to LTE-DC just overlap with very initial part of the LTE $T_{on}$ duration because the TXOP of LTE is larger than Wi-Fi. From the other side, LTE has a strong physical layer to compensate for any partial overlap/interference with it's subframe. So, in this calculation we assumed LTE-DC is collision free.

\section{ns-3 Simulation and Numerical Results}

ns-3 is an open source network simulator supporting credible simulation of protocol stacks for Wi-Fi and LTE \cite{nsnam}.
802.11a is well supported in ns-3 and for LTE-DC we obtained LTE signal with desired duty cycling using waveform generator function in ns-3. In all scenarios, both Wi-Fi and LTE-DC nodes are positioned within a small circle to ensure all nodes are exposed to each other. The full buffer traffic in UL/DL is considered and setup the flows of the network to be UDP in order to isolate the effect of transport layer on throughput. The  simulation code for an example coexistence scenario is available for download using the current version ns-3.27, see \cite{ns3example}. We consider two different data rates of 6 Mbps and 54 Mbps and two different $T_{C}$. As already noted, constant data rates only are considered (with no rate control). The other Wi-Fi network parameters are listed in Table~\ref{table: WiFipar}.

\begin{table}
\caption{Wi-Fi Parameters.}
\label{table: WiFipar}
\begin{center}
\begin{tabular}{|c|c|}
\hline 
Parameter & value \\
\hline
\hline
 PhyH & 20 $\mu s$ \\
 \hline
 MACH & (34 bytes)$/r_w$ $\mu s$ \\
 \hline
 $r_0$ & 6 Mbps, 12 Mbps, 24 Mbps \\
 \hline
 ACK & (14 bytes)$/r_0+20~\mu s$ \\
 \hline
 $\delta$ & 0.1 $\mu s$ \\
 \hline
 $\sigma$ & 9 $\mu s$ \\
 \hline
 DIFS & 34 $\mu s$ \\
  \hline
 SIFS & 16 $\mu s$ \\
  \hline
 $T_d$ & $N_B/r_w$ $\mu s$ \\
 \hline
\end{tabular}
\end{center}
\end{table}

\begin{figure*}[t]
\setlength{\belowcaptionskip}{-0.1in}
    \centering
    \begin{subfigure}{0.45\textwidth}
           \setlength{\belowcaptionskip}{-0.1in}
        \centering
        \centerline{\includegraphics[width=3.3in]{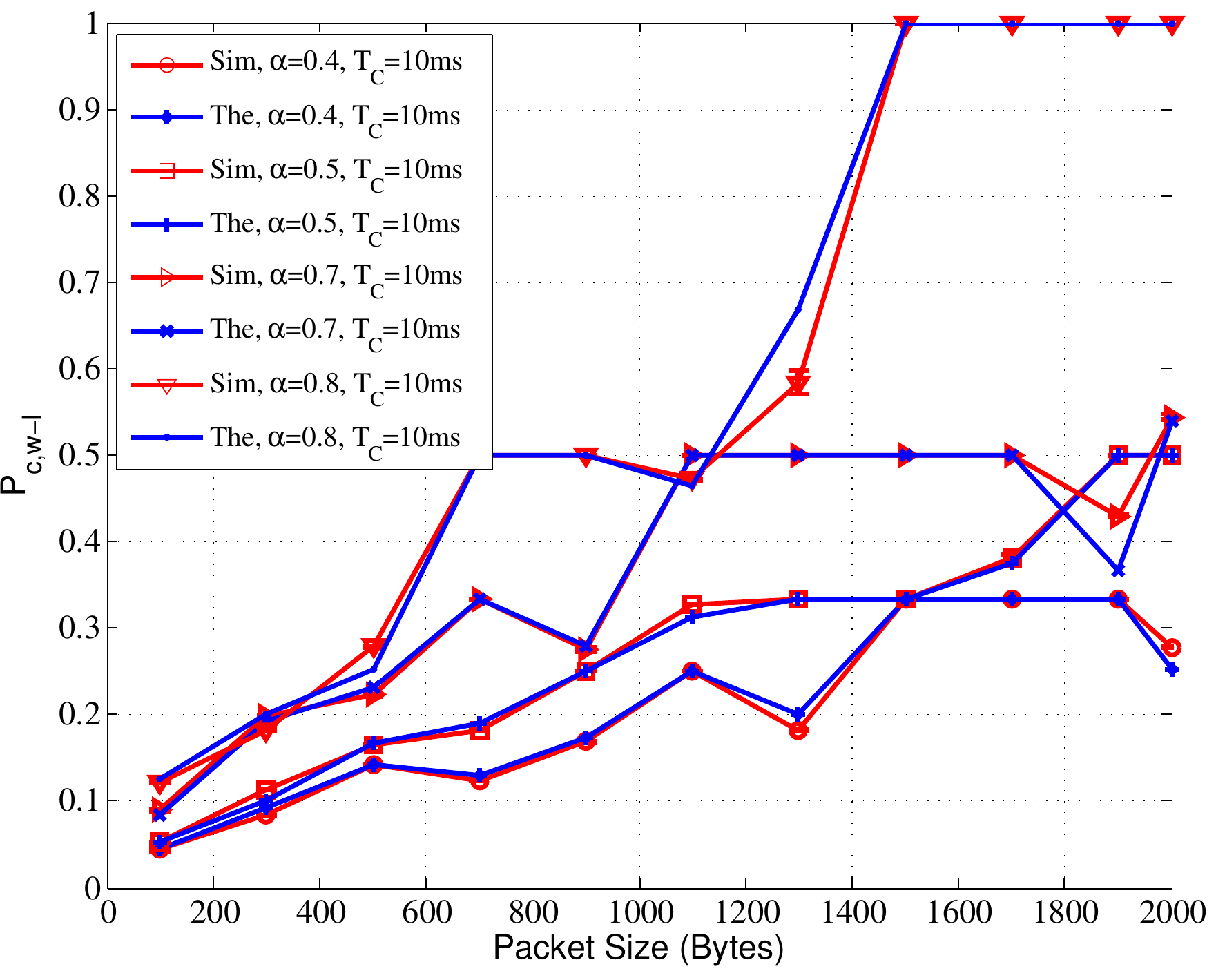}}
        \caption{Probability of Collision}
        \label{fig: Psim0}
    \end{subfigure}
    \hfill
    \begin{subfigure}{0.45\textwidth}
        \setlength{\belowcaptionskip}{-0.1in}
        \centering
        \centerline{\includegraphics[width=3.3in]{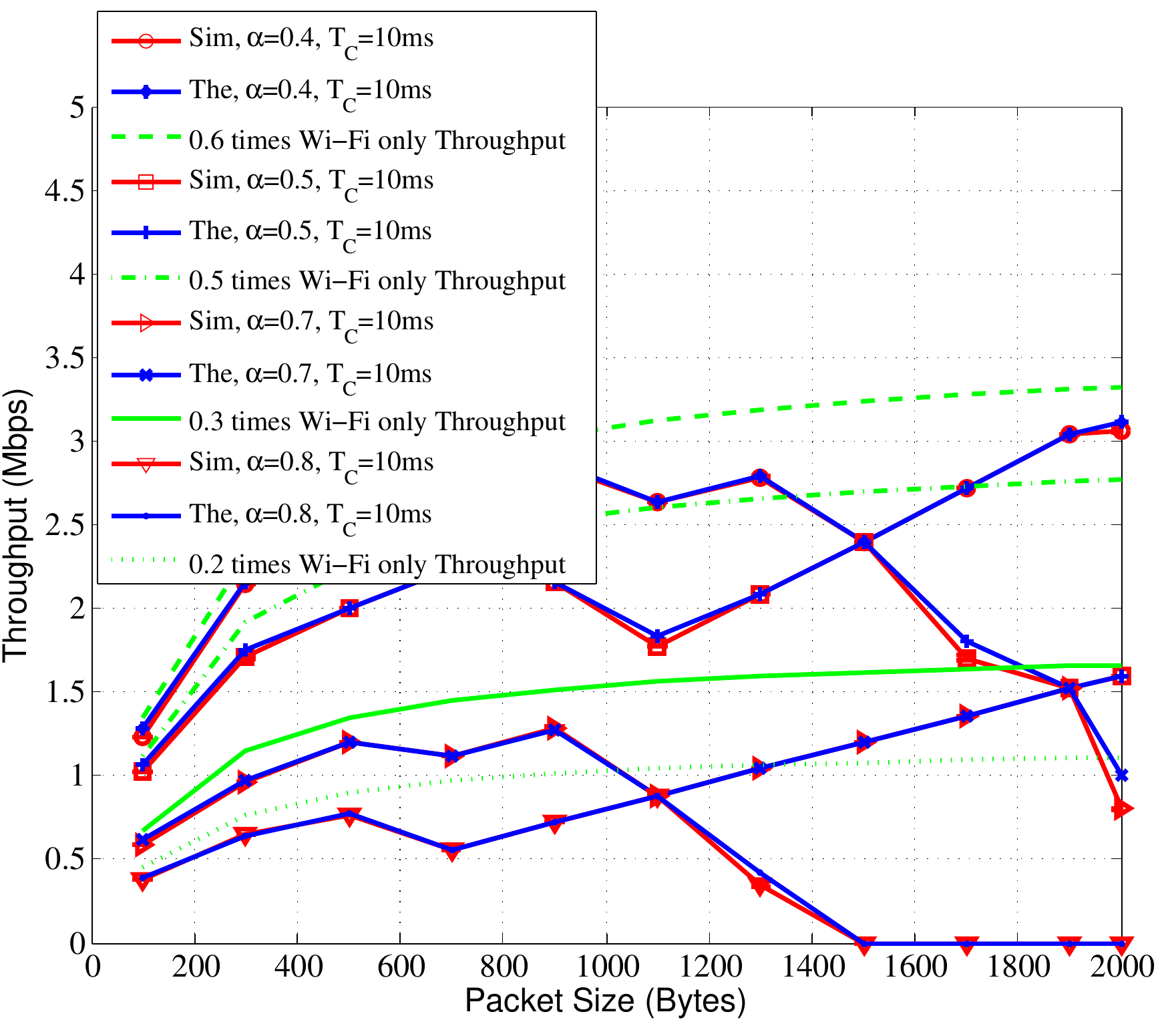}}
        \caption{Throughput}
        \label{fig: Tsim0}
    \end{subfigure}
    \caption{The comparison of theoretical and ns-3 simulation results for $T_{C}=10$ ms and Wi-Fi data rate of 6 Mbps. 1 Wi-Fi node (the AP transmitting in DL) coexisting with LTE-DC.}
    \label{fig: Sim0}
\end{figure*}

In Fig.~\ref{fig: Sim0} one Wi-Fi node (AP transmitting in DL using the data rate of 6 Mbps), coexists with LTE-DC operating with cycle period ($T_{C}$) of 10 ms, and duty cycle ($\alpha$) is varied between 0.4 to 0.7. Fig.~\ref{fig: Psim0} illustrates the probability of collision of Wi-Fi ($P_{c,w-l}$) in coexistence with LTE-DC. The theoretical estimates closely follow the simulation results. As can be seen, the collision probability fluctuates as a function of packet size because it depends on the packet airtime and the number of packets that fit into one OFF period. This means that the proportion of the random backoff duration and the packet airtime influences how many packets are transmitted through the fixed OFF period and affects the collision event of the final transmitted packet (with the ON edge). Increasing the duty cycle ($\alpha$) also increases the collision probability because fewer packets are transmitted during the OFF period and the last transmitted packet would collide with the ON edge, so the ratio of the collided to transmitted packet will increase. 

In Fig.~\ref{fig: Tsim0}, the simulated throughput performance of Wi-Fi AP in DL for duty cycles of 0.4 to 0.7 is shown to match theoretical model results. By increasing the duty cycle, the throughput decreases because fewer packets are transmitted and also the probability of collision is higher. The throughput fluctuation as a function of the packet size is seen in Fig.~\ref{fig: Psim0}, reflecting the variability in the packet collision probability mentioned above. The throughput of one Wi-Fi only system (with one Wi-Fi AP,  DL transmission) multiplied by the OFF cycle ($1-\alpha$) is shown in the figure as the upper bound of the Wi-Fi throughput if there is no interference from LTE-DC. Wi-Fi in coexistence with LTE-DC achieves a lower throughput than the Wi-Fi only case because of interference which leads to the collision of the last transmitted packet in OFF period. Generally, at smaller packet sizes the predicted throughput is closer to the upper bound, as more number of packets are transmitted in the OFF period, and hence the proportion of collided to transmitted packets decreases. We note that this fluctuation in the probability of collision and throughput was observed in \cite{Amr_LTEU} but presented without further analysis of underlying causes.

\begin{figure}[t]
\setlength{\belowcaptionskip}{-0.1in}
\centerline{\includegraphics[width=3.3in]{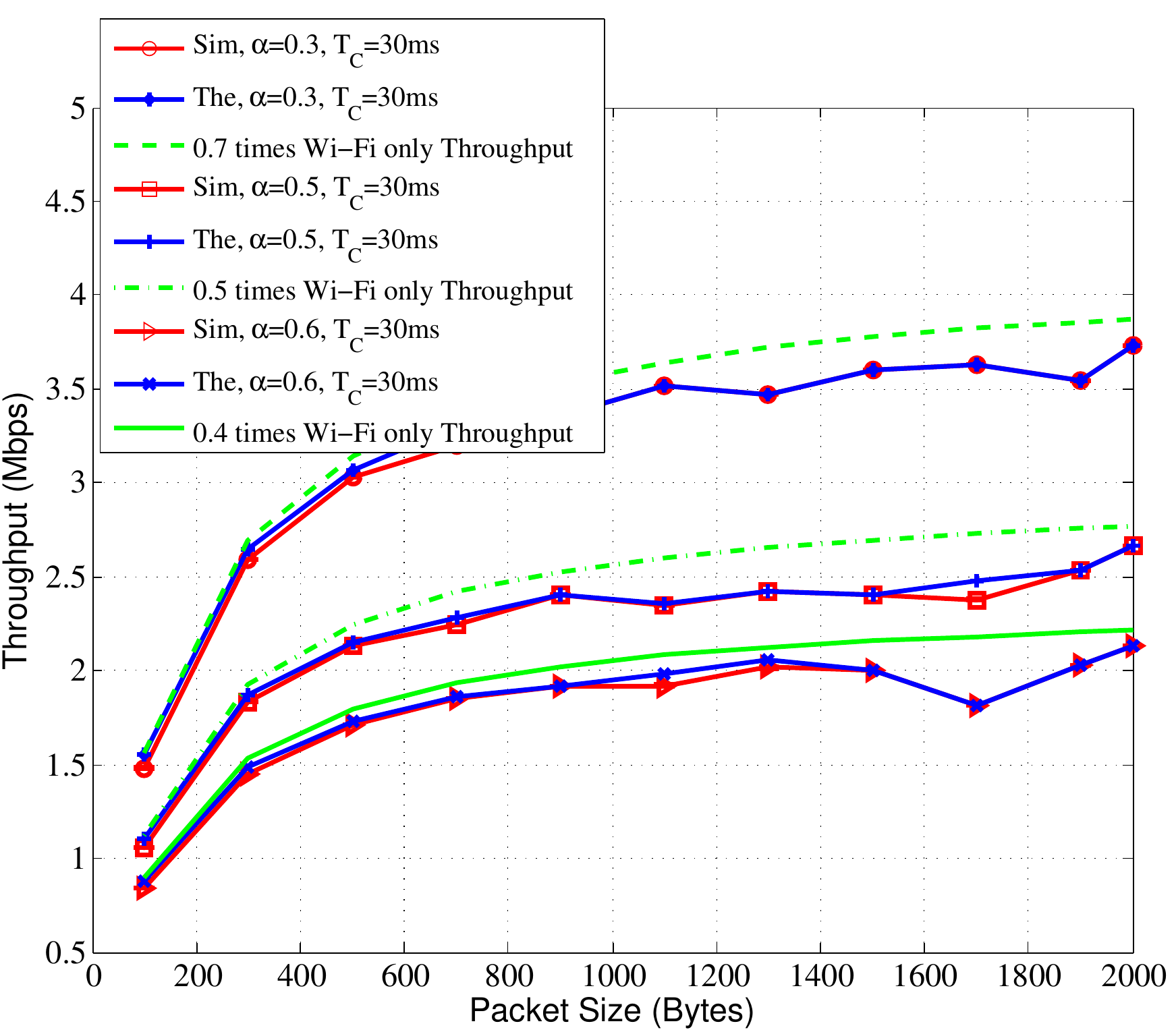}}
 \caption{The comparison of theoretical and ns-3 simulation results for $T_{C}=30$ ms and Wi-Fi data rate of 6 Mbps. 1 Wi-Fi node (the AP transmitting in DL) coexisting with LTE-DC.}
 \label{fig: Tsim1}
\end{figure}

In Fig.~\ref{fig: Tsim1}, the throughput of Wi-Fi AP in DL is illustrated for Wi-Fi data rate of 6 Mbps, cycle period $T_{C} = 30$ ms, and duty cycle range\footnote{Maximum duty cycle of 0.6 satisfies the maximum 20 ms LTE-DC ON period mandated by the standard.} 0.3 to 0.6 that shows very good agreement between our model estimates and ns-3 simulation.
Here, the throughput fluctuation is smaller compared with Fig.~\ref{fig: Tsim0} because the OFF period is larger. Moreover, the throughput performance of Wi-Fi in coexistence is closer to the duty cycled Wi-Fi only throughput. This also implies that the throughput, in this case, is higher than the throughput in Fig.~\ref{fig: Tsim0}. Therefore, to achieve a higher throughput at larger packet sizes, the $T_{C}$ should increase correspondingly. 

\begin{figure}[t]
\setlength{\belowcaptionskip}{-0.1in}
\centerline{\includegraphics[width=3.5in]{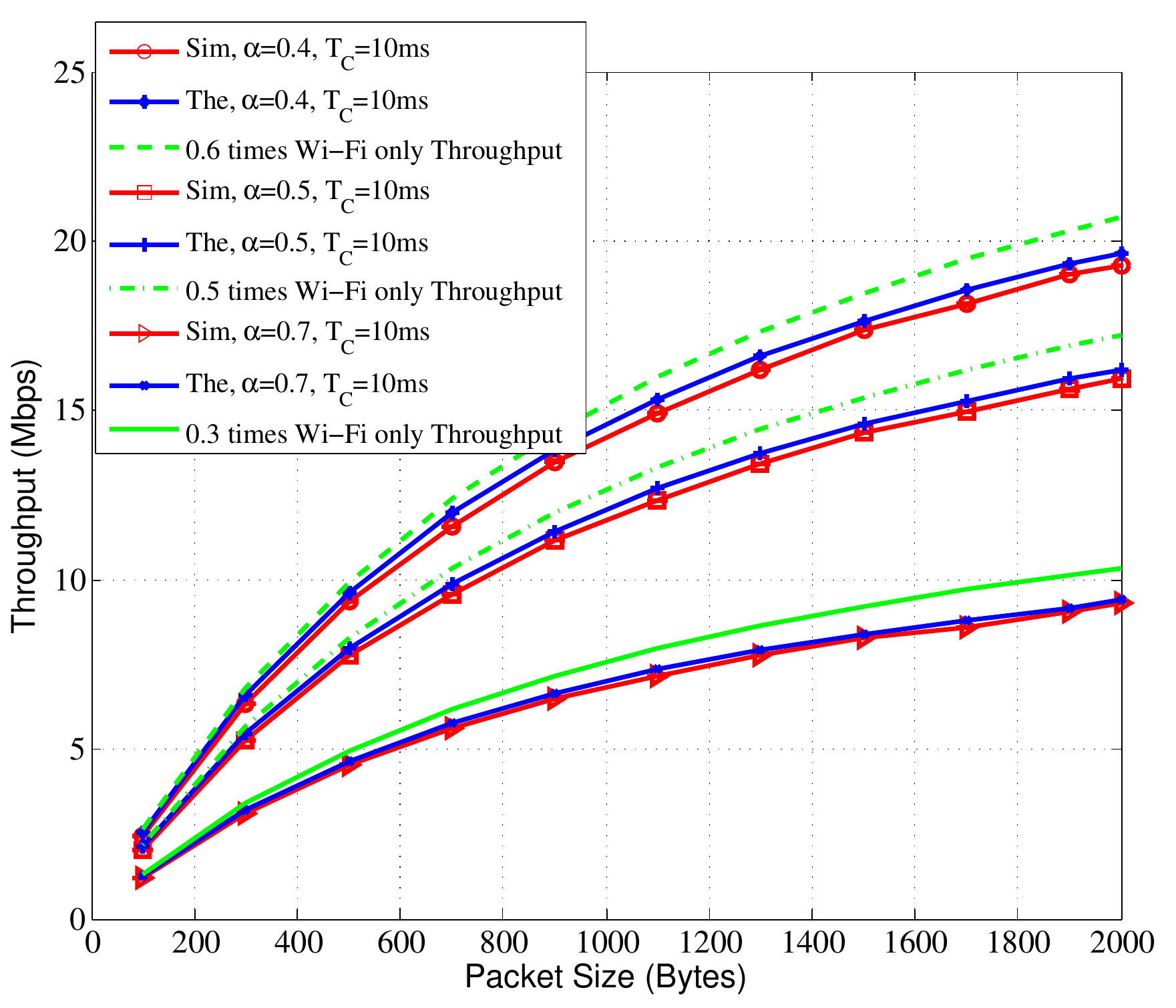}}
 \caption{The comparison of theoretical and ns-3 simulation results for $T_{C}=10$ ms and Wi-Fi data rate of 54 Mbps. 1 Wi-Fi node (the AP transmitting in DL) coexisting with LTE-DC.}
 \label{fig: Tsim2}
\end{figure}

In Fig.~\ref{fig: Tsim2}, the throughput performance of Wi-Fi AP in DL is illustrated where the Wi-Fi data rate is 54 Mbps, cycle period $T_{C} = 10$ ms, and duty cycles are 0.4 to 0.7. There is no fluctuation in the throughput performance in the plotted curves, due to smaller packet airtime at the higher data rate, which leads to fewer packet collisions. In this scenario, there is a small gap between the Wi-Fi throughput in coexistence system and Wi-Fi only throughput due to smaller interference from LTE-DC, i.e. a lower probability of collision.

\begin{figure}[t]
\setlength{\belowcaptionskip}{-0.1in}
    \centering
    \begin{subfigure}{0.45\textwidth}
           \setlength{\belowcaptionskip}{-0.1in}
        \centering
        \centerline{\includegraphics[width=3.5in]{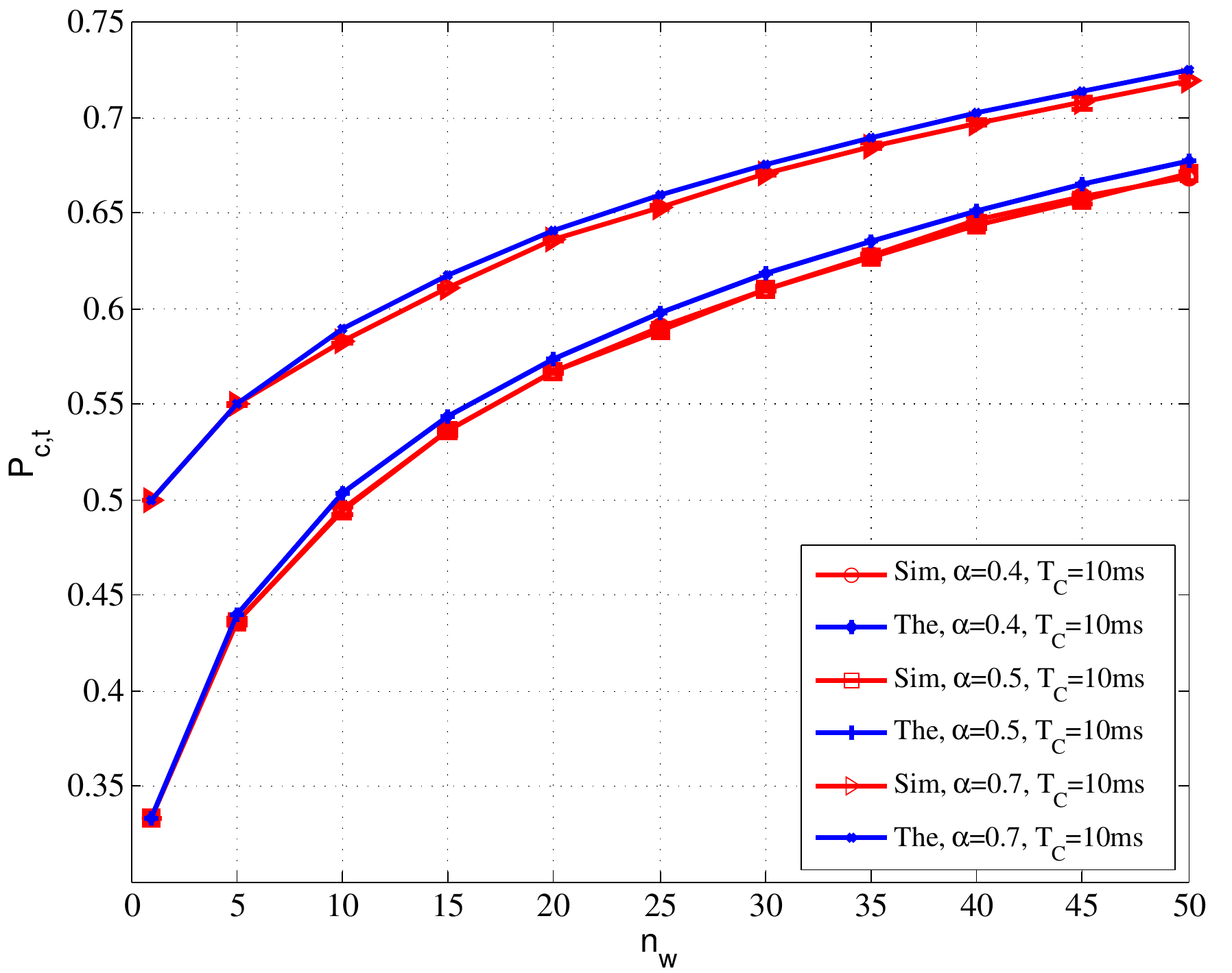}}
        \caption{Probability of Collision}
        \vspace{0.1in}
        \label{fig: Psim0n}
    \end{subfigure}
    \hfill
    \begin{subfigure}{0.45\textwidth}
        \setlength{\belowcaptionskip}{-0.1in}
        \centering
        \centerline{\includegraphics[width=3.5in]{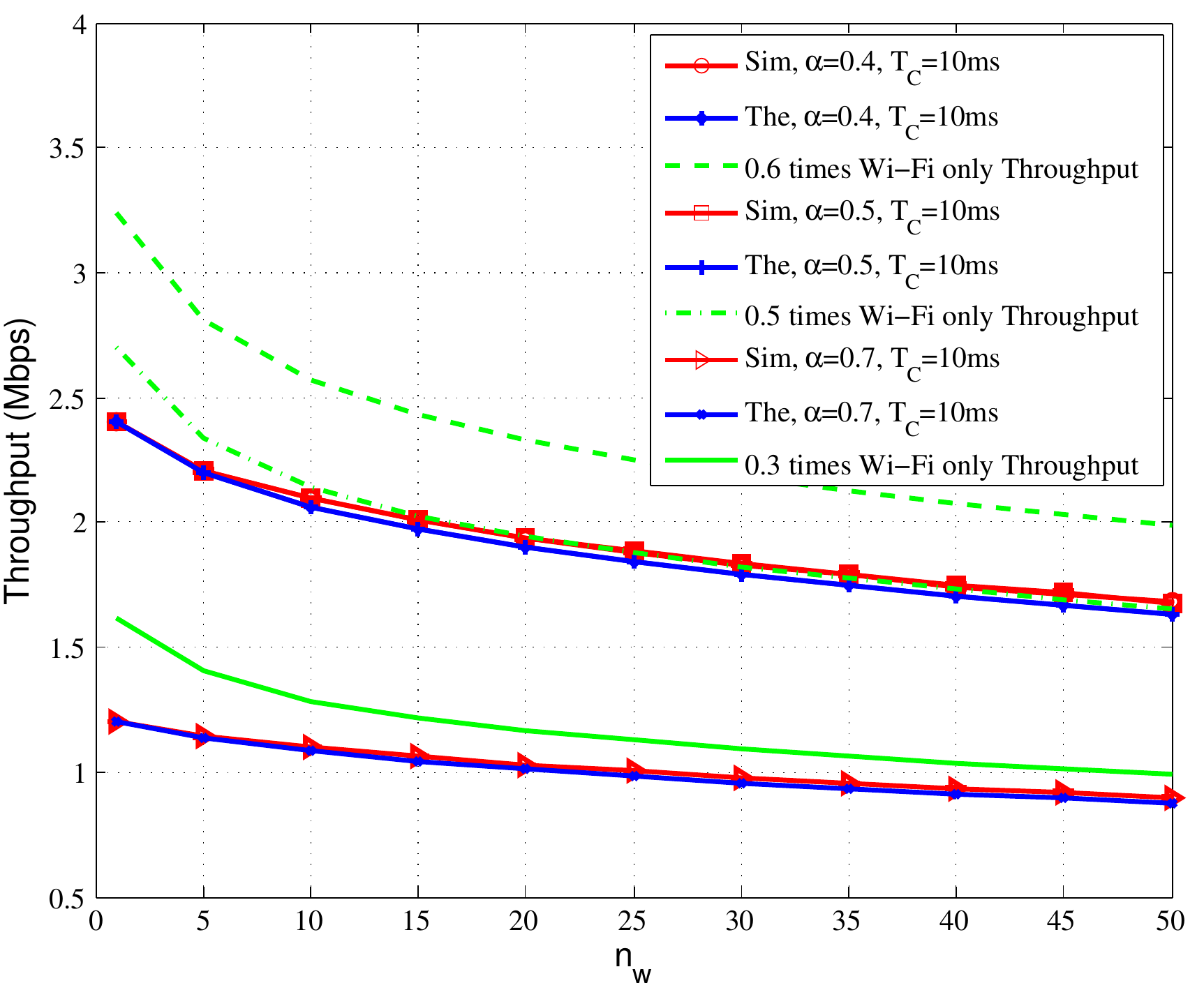}}
        \caption{Throughput}
        \label{fig: Tsim0n}
    \end{subfigure}
    \caption{The comparison of theoretical and ns-3 simulation results for $T_{C}=10$ ms and Wi-Fi data rate of 6 Mbps. $n_w$ Wi-Fi node transmitting in UL/DL and coexisting with LTE-DC.}
    \label{fig: Tsim3}
\end{figure}

In Fig.~\ref{fig: Tsim3}, $n_w$ Wi-Fi node (UL/DL transmission) coexists with LTE-DC with the Wi-Fi data rate of 6 Mbps, cycle period ($T_{C}$) of 10 ms, the packet size of 1500 bytes, and duty cycle ($\alpha$) is changing from 0.4 to 0.7. Fig.~\ref{fig: Psim0n} illustrates the probability of collision of Wi-Fi ($P_{c,t}$) and Fig.~\ref{fig: Tsim0n} the throughput of Wi-Fi in coexistence with LTE-DC via changing the number of stations. The theoretical results match the simulation results validating that the analytical model adequately captures  the inter-network collision probability in coexistence. The collision probability and throughput curves for 0.4 and 0.5 duty cycles are equal because, as can be seen in Fig.~\ref{fig: Tsim0} for one node, the collision probability and throughput for 1500 bytes are the same. This illustrates that depending on the packet airtime and LTE-DC OFF period, the same probability of collision and throughput can be achieved in coexistence for two different duty cycles (the same trend is valid for more number of nodes). The same happens at 1100 bytes for duty cycles of 0.7 and 0.8 in Fig.~\ref{fig: Tsim0}.

\begin{figure}[t]
\setlength{\belowcaptionskip}{-0.1in}
\centerline{\includegraphics[width=3.5in]{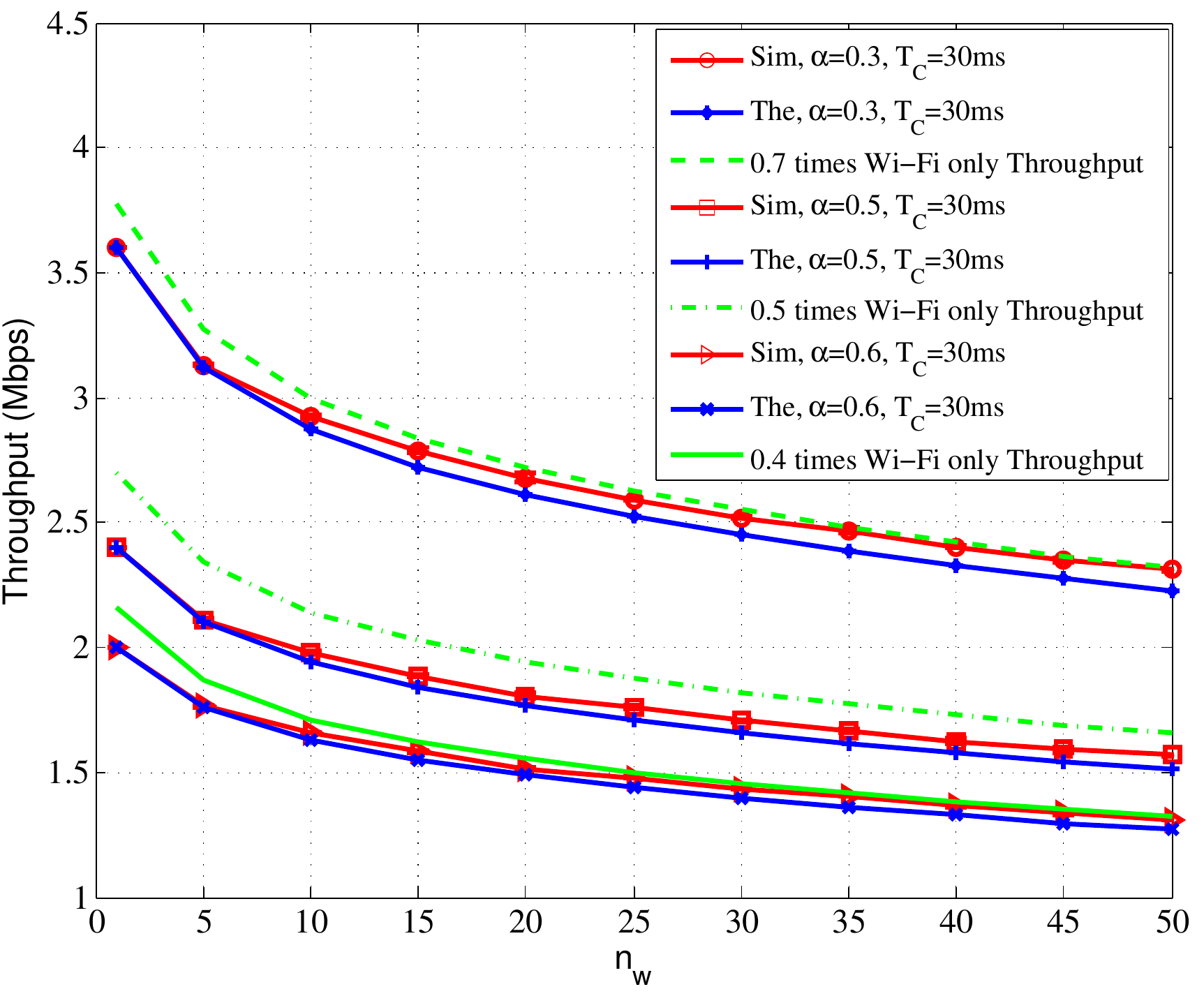}}
 \caption{The comparison of theoretical and ns-3 simulation results for $T_{C}=30$ ms and Wi-Fi data rate of 6 Mbps. $n_w$ Wi-Fi node transmitting in UL/DL and coexisting with LTE-DC.}
 \label{fig: Tsim4}
\end{figure}

\begin{figure}[t]
\setlength{\belowcaptionskip}{-0.1in}
\centerline{\includegraphics[width=3.5in]{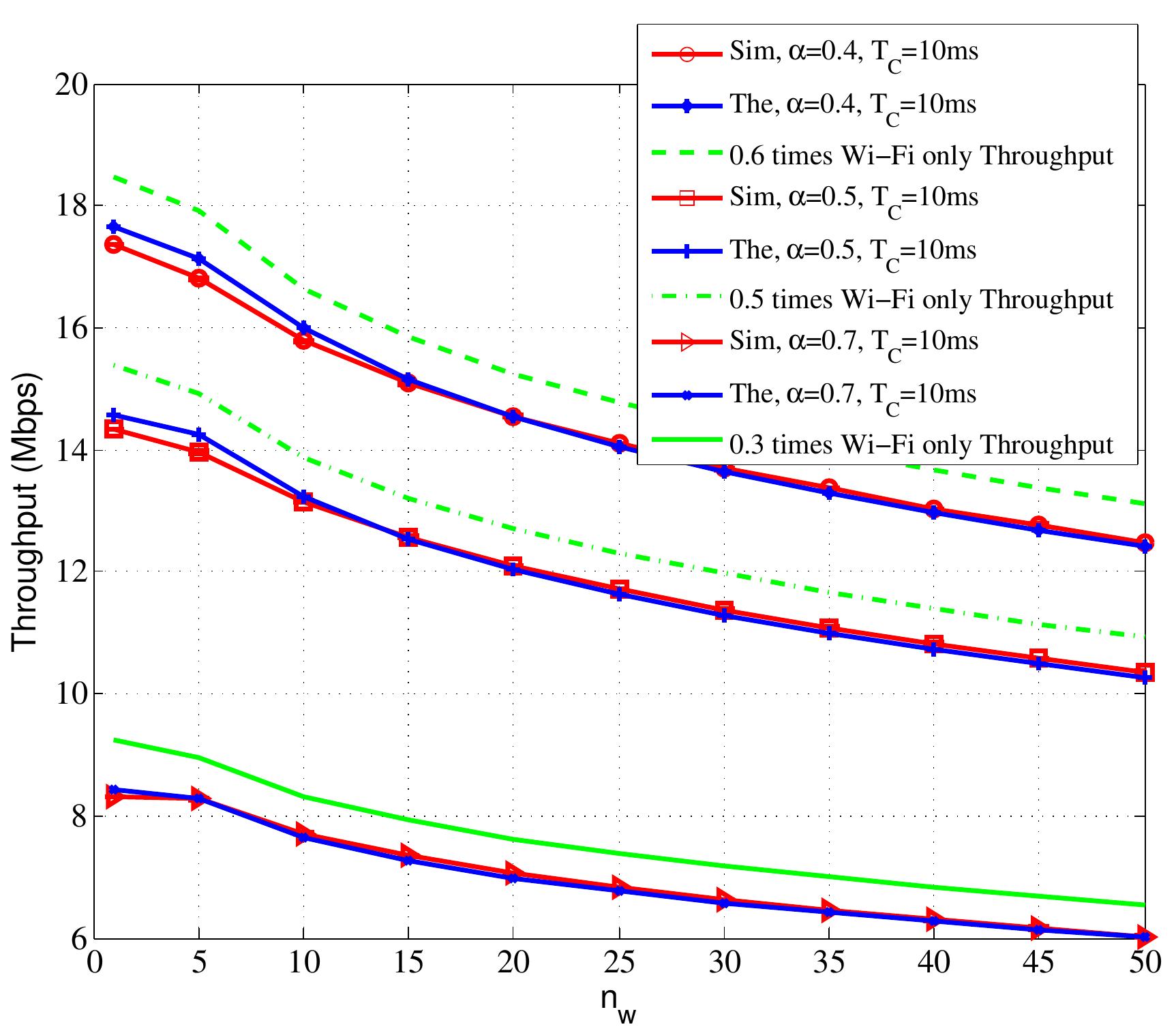}}
 \caption{The comparison of theoretical and ns-3 simulation results for $T_{C}=10$ ms and Wi-Fi data rate of 54 Mbps. $n_w$ Wi-Fi node transmitting in UL/DL and coexisting with LTE-DC.}
 \label{fig: Tsim5}
\end{figure}

In Fig.~\ref{fig: Tsim4} the scenario is the same as Fig.~\ref{fig: Tsim3}, except the cycle period, is 30 ms, and the duty cycle ($\alpha$) is varied from 0.3 - 0.6. For the duty cycle of 0.3 and 0.6, the Wi-Fi throughput is closer to the Wi-Fi only throughput because the proportion of packet airtime and OFF period is such that the number of packets that can be successfully transmitted during OFF period is larger. The scenario in Fig.~\ref{fig: Tsim5} is the same as Fig.~\ref{fig: Tsim4} except the cycle period is 10 ms, the data rate is 54 Mbps, and the duty cycle is varied from 0.4 - 0.7. The theoretical and simulation results are suitably matching. There is a gap between the Wi-Fi throughput in coexistence and Wi-Fi only throughput because of collision with the LTE-DC ON edge. The gap is almost the same for different duty cycles; this is due to the small packet airtime compared with the OFF period which leads to a small and almost equal probability of collision which leads to throughput loss of Wi-Fi in coexistence system compared with Wi-Fi only system (gap between the curves). 

\textbf{Summary:}
\begin{itemize}
    \item The fluctuation of throughput for 6 Mbps and $T_{C}=10$ ms scenario results from the packet airtime being comparable with $T_{off}$; throughput curves are smooth for the other two scenarios, i.e. 6 Mbps in 30 ms and 54 Mbps in 10 ms. However, the same behavior for higher data rate, e.g. 54 Mbps, is expected when a station is allowed to transmit over TXOP period in which the TXOP is comparable with $T_{off}$ in IEEE 802.11n and 802.11ac standards. The effect of TXOP and block ACK in coexistence with LTE-DC is deferred for future work. 
    
    \item When the $T_{off}$ is much larger than the packet airtime, the Wi-Fi only throughput multiplied by the duty cycle ($\alpha$) is a good approximation of Wi-Fi throughput in coexistence network as can be seen in Fig.~\ref{fig: Tsim2} (especially at smaller packet sizes and it deviates for larger packet sizes). 
\end{itemize}


\section{Fairness in Coexistence}

A pragmatic approach to `fair' Wi-Fi and LTE-DC coexistence is to achieve a suitably defined fairness metric by tuning the LTE-DC parameters $T_{C}$ and $\alpha$.  While the LTE-U Forum did not provide any formal definition and test scenarios of `fair coexistence',  there have been several investigations on this topic - notably \cite{Qualcomm} and \cite{Cablelabs2} whose conclusions trend in opposite directions. In the investigation by CableLabs \cite{Cablelabs2}, they first established a baseline of how two Wi-Fi networks share a channel; then, they replaced one of the Wi-Fi APs with an LTE-U eNB and repeated the tests, using various duty cycle configurations. This parallels the 3GPP definition of fairness for the LTE-LAA coexistence with Wi-Fi \cite{3GPP_TR} whereby ``LAA design should target fair coexistence with existing Wi-Fi networks, so as to not impact Wi-Fi services more than an additional Wi-Fi network on the same carrier, with respect to throughput and latency''. Hence, we use this notion of fairness for our own investigations below. 

With reference to the operating scenario in Fig.~\ref{fig: Diag}, we say that fairness is achieved if the appropriate metric (access or throughput) is not worse if a Wi-Fi network  is replaced by an LTE-DC network, i.e. the same access probability or throughput is achieved for scenario in Fig.~\ref{fig: Diag} (a) and Fig.~\ref{fig: Diag} (b). We assume that in scenario (a), there are $n_w$ stations in each Wi-Fi network, so totally $N=2n_w$ Wi-Fi stations are contending for transmission; in scenario (b) the $n_w$ Wi-Fi stations are replaced with an LTE-DC network in which the eNB is transmitting on downlink to the stations.

\subsection{Wi-Fi Access Fairness}

Access fairness is achieved by tuning the $\alpha$ or $T_{C}$ such that the probability of channel access is identical in both scenarios. In Wi-Fi only scenario with two networks and total $N=2n_w$ Wi-Fi stations (twice as the number of Wi-Fi stations in the coexistence scenario) - Fig.~\ref{fig: Diag} (a) - the probability of channel access is 
\begin{equation}
\begin{split}
& \tau_{wo} =\\ & \frac{2}{W_0\left(\frac{(1-(2P_{c,wo})^{m+1})(1-P_{c,wo})+2^{m}\left(P_{c,wo}^{m+1}-P_{c,wo}^{m+2}\right)(1-2P_{c,wo})}{(1-2P_{c,wo})(1-P_{c,wo}^{m+2})}\right) +1},\\
&P_{c,wo} = 1-(1-\tau_{c,wo})^{N-1},
\label{eq: tauwo}
\end{split}
\end{equation}
where using (\ref{eq: tauw}) in which $P_{c,wo}$ is replaced by $P_{c,w}$. 

In the coexistence of Wi-Fi and LTE-DC, just $n_w$ Wi-Fi stations, belonging to one Wi-Fi network, contend for channel access during the LTE-DC OFF period. So, as calculated in the previous section, the probability of accessing the channel is:
\begin{equation}
\begin{split}
&\tau_w(\alpha,T_{C}) =\\ &\frac{2}{W_0\left(\frac{(1-(2P_{c,t})^{m+1})(1-{P_{c,t}})+2^{m}\left({P_{c,t}}^{m+1}-{P_{c,t}}^{m+2}\right)(1-2P_{c,t})}{(1-2P_{c,t})(1-{P_{c,t}}^{m+2})}\right) +1},\\
&P_{c,t}(\alpha,T_{C}) = 1-(1-P_{c,w})(1-P_{c,w-l}),
\label{eq: tauwc}
\end{split}
\end{equation}
where $\tau_w$ is a function of $\alpha$ and $T_{C}$. This follows from (\ref{eq: tauw}) with $P_{c,w}$ replaced by the $P_{c,t}$.  To achieve access fairness, we thus propose to
\begin{equation}
\begin{split}
&\underset{\alpha}{\min}~ \left | \tau_{w}(\alpha,T_{C}) - \tau_{wo} \right|,\\
& s.t.~~0 < \alpha < 1,
\end{split}
\label{eq: Opt1}
\end{equation}
where the $T_{C}$ range depend on the maximum and minimum values of ON and OFF period of LTE-DC, per the LTE-U standard.

\subsection{Wi-Fi Throughput Fairness}

In coexistence, the Wi-Fi throughput (\ref{eq: tputwc}) depends on the LTE-DC parameters $\alpha$ and $T_{C}$, which can be tuned to achieve Wi-Fi throughput. For a given $T_{C}$, we optimize $\alpha$ to achieve throughput fairness. 
For the two Wi-Fi networks with $N=2n_w$ contending Wi-Fi stations, the total throughput in Wi-Fi only is calculated as:
\begin{equation}
\begin{split}
&Tput_{wo} =\frac{P_{tr} P_{s} T_d}{(1-P_{tr})\sigma+P_{tr}(1-P_{s})T_{cw}+P_{tr}P_{s}T_{sw}}r_w, \\
&P_{tr} =1-(1-\tau_{wo})^{N},\\
&P_{s} =\frac{N \tau_{wo} (1-\tau_{wo})^{N-1}}{P_{tr}},\\
&T_{sw} = \text{MACH}+\text{PhyH}+T_d+\text{SIFS}+\delta+\text{ACK}+\text{DIFS}+\delta \\
&T_{cw} = T_{sw}
\label{eq: tputwo}
\end{split}
\end{equation}
where the $\tau_{wo}$ and collision probability $P_{wo}$ is given by Eq (\ref{eq: tauwo}). So, in the coexistence network $n_w$ Wi-Fi stations, belonging to one Wi-Fi network, contend for channel access during the LTE-DC OFF period. We propose to tune $\alpha$ of LTE-DC such that the throughput fairness is achieved as follows:
\begin{equation}
Tput_{w}(\alpha,T_{C}) = \frac{Tput_{wo}}{2},
\label{eq: wfair}
\end{equation}
i.e., the optimization problem to achieve Wi-Fi throughput fairness is 
\begin{equation}
\begin{split}
& \underset{\alpha}{\min}~ \left | Tput_{w}(\alpha,T_{C}) - \frac{Tput_{wo}}{2} \right|\\
& s.t.~~0 < \alpha < 1.
\end{split}
\label{eq: Opt2}
\end{equation}

\section{ Coexistence Fairness: Numerical Results}

The two different notions of fairness in coexistence of Wi-Fi and LTE-DC network are investigated for three different scenarios: A) $T_{C}=10$ ms with $r_w=6$ Mbps, B) $T_{C}=30$ ms with $r_w=6$ Mbps, and C) $T_{C}=10$ ms with $r_w=54$ Mbps. The packet size is kept fixed at $1500$ bytes and the LTE duty cycle parameter $\alpha$ is tracked as a function of the number of stations to achieve the desired fairness metric. The fairness problems in eq. (\ref{eq: Opt1}) and (\ref{eq: Opt2}) is numerically solved using the MATLAB numerical solver. Since we already validated the analytical model, we are just using the numerical results for investigating the fairness. 

\subsection{Wi-Fi Access Fairness}

\begin{figure}[t]
\setlength{\belowcaptionskip}{-0.1in}
\centerline{\includegraphics[width=3.5in]{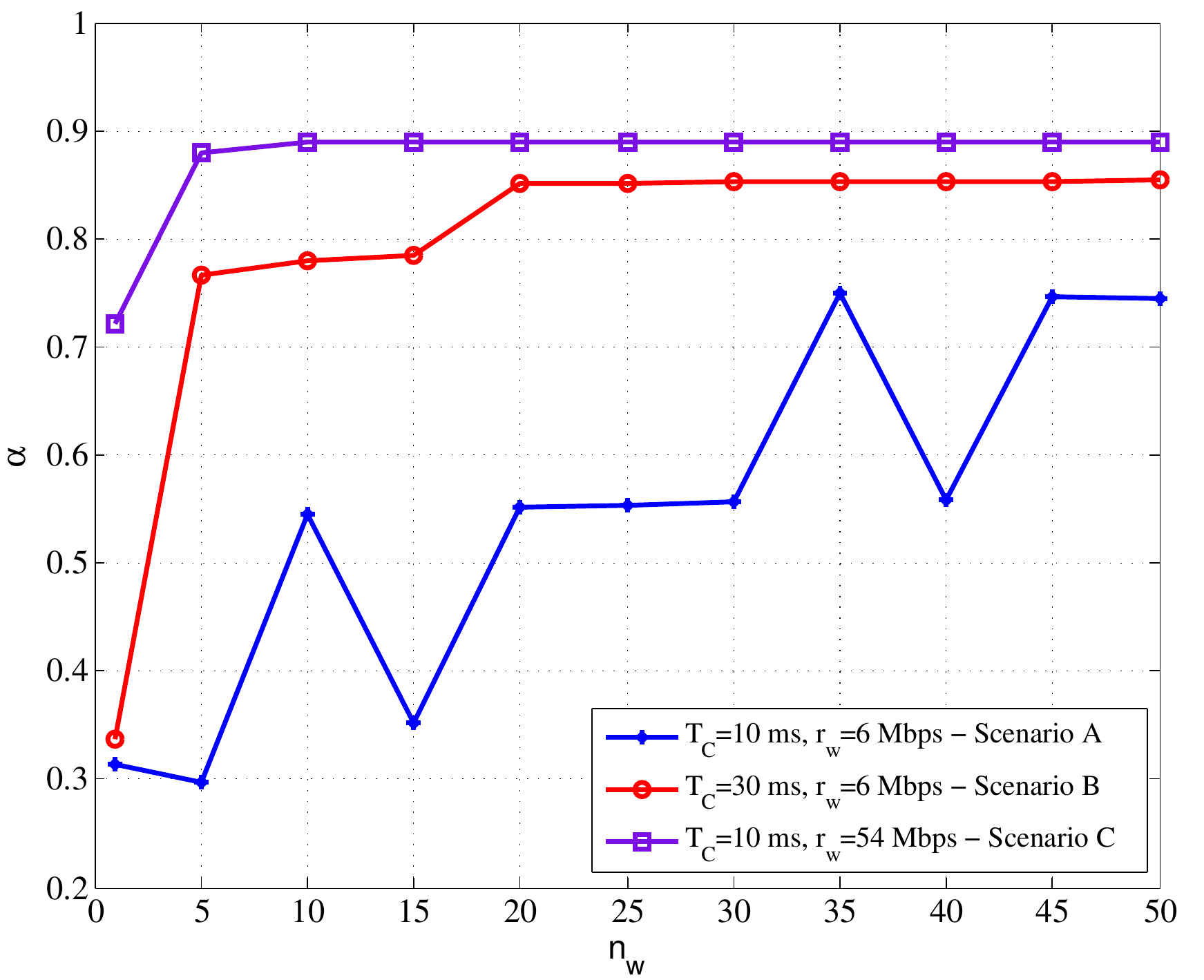}}
 \caption{The optimized duty cycle ($\alpha$) to achieve the access fairness vs. different number of stations - the Wi-Fi packet size is 1500 bytes.}
 \label{fig: AccessFair}
\end{figure}

In access fairness, the LTE duty cycle ($\alpha$) is chosen such that the same access probability of Wi-Fi in coexistence and Wi-Fi only network is achieved. Fig.~\ref{fig: AccessFair} illustrates the optimized $\alpha$ versus the number of Wi-Fi stations, for the three different scenarios. In all of these scenarios, $\alpha$ could be very large and the LTE-DC gets most of the airtime. So, the available airtime and correspondingly the throughput of Wi-Fi is very small. This fact implies that by achieving the access fairness, the LTE-DC is very unfair to Wi-Fi regarding airtime and throughput. 

\subsection{Wi-Fi Throughput Fairness}

\begin{figure}[t]
\setlength{\belowcaptionskip}{-0.1in}
\centerline{\includegraphics[width=3.5in]{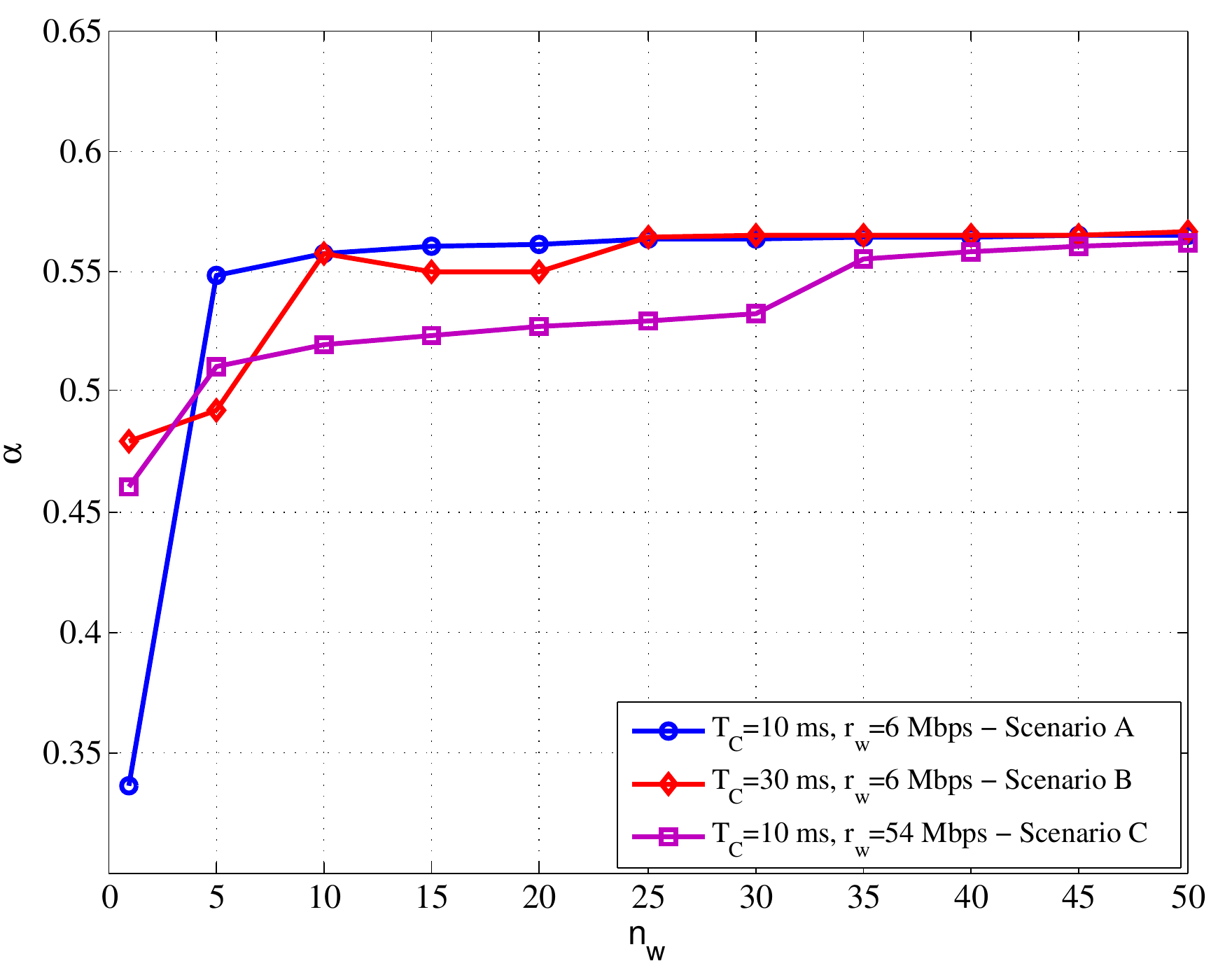}}
 \caption{The optimized duty cycle ($\alpha$) to achieve the throughput fairness vs. different number of stations - the Wi-Fi packet size is 1500 bytes.}
 \label{fig: TputFair}
\end{figure}

\begin{figure}[t]
\setlength{\belowcaptionskip}{-0.1in}
\centerline{\includegraphics[width=3.5in]{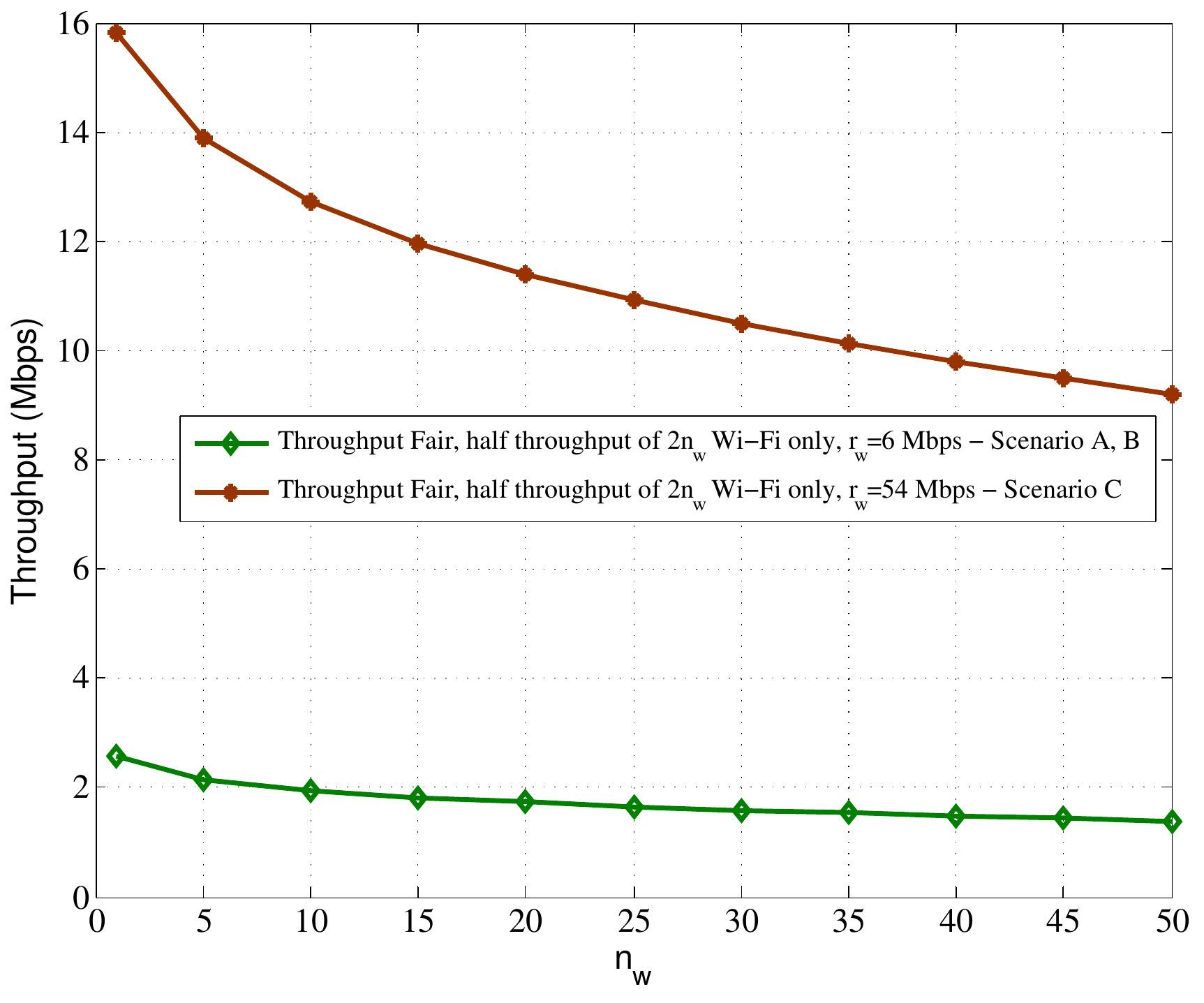}}
 \caption{The Wi-Fi throughput (half throughput, i.e. $\frac{Tput_{wo}}{2}$, which is calculated from (\ref{eq: tputwo})) when the throughput fairness is achieved for the Wi-Fi packet size of 1500 bytes.}
 \label{fig: TWiFiFair}
\end{figure}

Fig.~\ref{fig: TputFair} shows the optimized $\alpha$ for throughput fairness for same scenarios as Fig.~\ref{fig: AccessFair}. For one Wi-Fi client and downlink, scenario A requires the smallest LTE-DC duty cycle for achieving the fairness, while for scenario B, $\alpha$ is the largest. This is because the packet airtime is comparable with the LTE-DC OFF duration, i.e. the OFF duration is very small in scenario A. As the number of stations increases, the duty cycle increases which implies that Wi-Fi - Wi-Fi coexistence causes more throughput drop than the LTE-DC. Hence, LTE-DC can use a larger portion of the airtime and still be fair to Wi-Fi compared with another Wi-Fi. Fig.~\ref{fig: TWiFiFair} shows the throughput of Wi-Fi when fairness is achieved - note that the curve with $2n_w$ node and $r_w=6$ Mbps is the throughput for scenario A and B, and the curve with $2n_w$ node and $r_w=54$ Mbps is the throughput for scenario C.

\textbf{Summary:}
\begin{itemize}
    \item Access fairness results in a very large optimal LTE-DC duty cycle which implies a very small throughput for Wi-Fi.
    
    \item In throughput fairness, as the number of Wi-Fi nodes increases, the Wi-Fi network requires smaller airtime (larger $\alpha$) to achieve fairness. This implies that for duty cycle value $\alpha=0.5$, Wi-Fi could achieve a higher throughput in coexistence with LTE-DC than another similar size Wi-Fi network.
    
    \item The LTE-U Forum specifications \cite{LTEU_V13} recommend that LTE-U share the channel with one full buffer Wi-Fi link (one Wi-Fi network) by tuning its duty cycle below 0.5. For $n_w = 1$, our results show that throughput fairness satisfies the LTE-U Forum condition but for $n \ge 5$, this condition is not satisfied. 
    
\end{itemize}

The above results are taken together suggest that the fairness metrics considered may not be sufficient; thus, a deeper investigation of this matter is necessary and is deferred to future work.

\section{Conclusion}

In this work, we first presented a new analytical model for computing the throughput performance of Wi-Fi in coexistence with LTE-DC. The analytical results suitably match the ns-3 simulation results, validating the proposed model. Further, access and throughput fairness of Wi-Fi in coexistence with the LTE-DC network are investigated as a function of the LTE-DC duty cycle. The results indicate that in most of the scenarios considered: by increasing the number of Wi-Fi nodes, the Wi-Fi network in coexistence with LTE-DC with $\alpha=0.5$ achieves a higher throughput than coexisting with another similar Wi-Fi network. Several directions of future work are indicated: an analytical model for LTE-U with CSAT, and a deeper investigation into coexistence fairness between Wi-Fi and LTE-U.

\section*{ACKNOWLEDGMENT}
The authors would like to thank Drs. Monisha Ghosh and Vanlin Sathya (U Chicago) for helpful discussions.

\bibliographystyle{IEEEtran}
\bibliography{WiFiCoex}

\end{document}